\documentclass[11pt,a4paper]{article}
\usepackage{cmap}
\usepackage[T1]{fontenc}
\usepackage[utf8]{inputenc}
\usepackage{lmodern}
\usepackage{textcomp}
\usepackage{microtype}
\DisableLigatures{encoding = *, family = *}
\input{glyphtounicode.tex}
\usepackage{amsmath,amssymb}
\usepackage{graphicx}
\usepackage{booktabs}
\usepackage[margin=1in]{geometry}
\usepackage{natbib}
\usepackage{caption}
\usepackage{subcaption}

\usepackage{enumitem}
\usepackage{hyperref}

\title{HHL with a Coherent Fourier Oracle: A Proof-of-Concept\\
Quantum Architecture for Joint Melody-Harmony Generation}

\author{Alexis Kirke}
\date{}

\begin{document}
\maketitle

\begin{abstract}
Quantum algorithms with a proven theoretical speedup over classical computation
are rare.  Among the most prominent is the Harrow-Hassidim-Lloyd (HHL)
algorithm for solving sparse linear systems.  Here, HHL is applied to encode
melodic preference: the system matrix encodes Narmour implication-realisation
and Krumhansl-Kessler tonal stability, so its solution vector is a
music-cognition-weighted note-pair distribution.  The key constraint of HHL
is that reading its output classically cancels the quantum speedup; the
solution must be consumed coherently.  This motivates a \emph{coherent
Fourier harmonic oracle}: a unitary that applies chord-transition weights
directly to the HHL amplitude vector, so that a single measurement jointly
selects both melody notes and a two-chord progression.

A two-note/two-chord (2/2) block is used to contain the exponential growth
of the joint state space that would otherwise make classical simulation of
larger blocks infeasible.  For demonstrations of longer passages, blocks are
chained classically - each block's collapsed output conditions the next
-- as a temporary workaround until fault-tolerant hardware permits larger
monolithic circuits.  A four-block chain produces 8 notes over 8 chords with
grammatically valid transitions at every block boundary.

Independent rule-based harmony validation confirms that 97\% of generated
chord progressions are rated strong or acceptable.  The primary motivation
is that HHL carries a proven exponential speedup over classical linear
solvers; this work demonstrates that a coherent HHL+oracle pipeline - the
prerequisite for that speedup to be realised in a musical setting - is
mechanically achievable.
Audio realisations of representative outputs are available for listening at
\url{https://alexiskirke.com/quantum-hhl-music-results-files/}.
\end{abstract}

\section{Introduction}

Quantum algorithms with a proven theoretical speedup over classical computation
are rare.  Among the most prominent are Shor's algorithm for integer
factorisation~\citep{shor1994algorithms}, Grover's algorithm for unstructured
search~\citep{grover1996fast}, and the Harrow-Hassidim-Lloyd (HHL) algorithm
for solving sparse linear systems~\citep{harrow2009quantum}.  A research
programme has been underway to apply these algorithms to computer music,
asking whether algorithms with genuine quantum advantage can be meaningfully
connected to musical problems.  Earlier work applied Grover's algorithm to
rule-based melodic composition on IBM quantum
hardware~\citep{kirke2019applying}, and then extended it to multi-agent
interactive performance via quantum teleportation~\citep{kirke2020testing}.
This paper presents what appears to be the first application of HHL as
a functional music generation engine - as a preference-driven compositional
architecture rather than as a tool for classical post-processing or
sonification~\citep{weimer2010quantenblog}.

A recurring question in algorithmic composition research is the coupling of
melody and harmony: how do they inform each other?  Constraint-based systems
handle this by solving a joint satisfaction problem
\citep{truchet2011constraint, anders2017constraint}.  Sequential systems choose
one first and condition the other on it.  Both approaches assume discrete,
observable steps.

HHL naturally suggests a different approach.  Applying HHL correctly requires
that its solution vector not be read out classically - classical readout would
eliminate the speedup.  Instead the solution must be consumed coherently by a
downstream quantum process.  In quantum computing, an \emph{oracle} is a
black-box subroutine that encodes a rule or function directly into the
circuit - think of it as the quantum equivalent of a look-up table or
filter that a classical algorithm would call as a function.  For a computer
musician, the oracle here is the harmonic rule-book: given the melody
amplitudes produced by HHL, it weights each possible chord according to how
well it fits, without ever stopping to read out an intermediate result.  In
this architecture, that process is a
\emph{coherent harmonic oracle}: a unitary operation that applies harmonic
preferences directly to the HHL amplitude vector without collapsing the melodic
superposition.  Melody and harmony are then resolved together coherently.  Crucially, avoiding collapse between HHL and the oracle is what preserves the
conditions under which an HHL speedup could apply at scale; it is not claimed
to improve the musical output at the present simulation scale, where the
results are statistically indistinguishable from a classical Markov chain.

\begin{figure}[h]
  \centering
  \includegraphics[width=\textwidth]{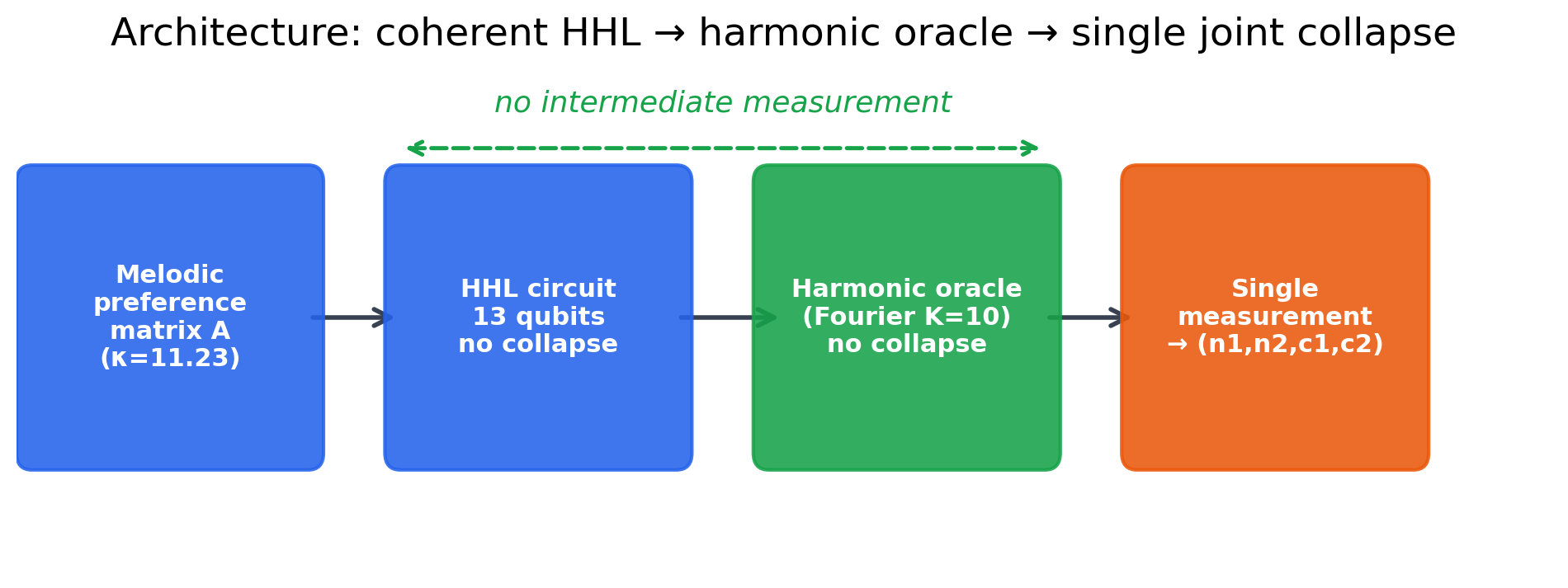}
  \caption{Coherent HHL $\to$ Fourier harmonic oracle pipeline: no intermediate measurement.
           A single joint collapse selects melody notes and two-chord progression simultaneously.}
  \label{fig:arch}
\end{figure}

This paper is a proof-of-concept study in the same spirit
as~\citet{kirke2019applying} and~\citet{kirke2020testing}: the scale is
deliberately minimal (49 note pairs, 7 chord functions, 19 qubits), and the
goal is to demonstrate that HHL can be applied to music generation in a
quantum-coherent manner.  HHL's exponential speedup over classical linear
solvers cannot be accessed without a coherent end-to-end pipeline; the
contributions below are the steps toward establishing that pipeline.
Three technical contributions are made beyond the
baseline HHL+oracle architecture: (i) the oracle is implemented without
per-pair normalisation of the chord register, preserving global amplitude
coherence; (ii) the Fourier oracle is realised as a concrete Qiskit circuit
and characterised empirically; and
(iii) a four-block 2/2 chaining architecture with melodic and harmonic context
conditioning is demonstrated.

\section{Related Work}

Quantum computer music (QCM) research can be usefully partitioned along two
axes: whether the underlying quantum algorithm carries a mathematically proven
speedup over the best classical equivalent, and whether the algorithm is
deployed to solve a musical problem functionally or merely to sonify the
algorithm's own behaviour.  This yields three types.  \emph{Type~1}: proven
speedup, used functionally to generate music - the rarest category.  \emph{Type~2}: no proven
speedup, used for sonification of algorithmic behaviour.  \emph{Type~3}: no
proven speedup, used functionally for music generation (variational circuits,
quantum annealing, cellular automata).  The present paper is Type~1.
We are not claiming Type~1 work is superior, but merely that there is a clear research gap in this area.

The origins of quantum computer music trace to early informal experimentation,
most notably an online blog post exploring the idea of quantum musical
generation~\citep{weimer2010quantenblog}, before academic work began in earnest.
Systematic research started with Kirke's papers on photonic quantum systems
and hybrid quantum-classical architectures, including a photonic quantum
computer simulation~\citep{kirke2016hybrid} and the Q-Muse system performed
live with a photonic quantum computer~\citep{kirke2014qmuse}.
Further work explored hybrid quantum-classical frameworks for interactive
performance~\citep{kirke2017experiments}.
A significant step came with the application of Grover's algorithm to
rule-based melodic composition on IBM quantum hardware
\citep{kirke2019applying} - the first use in music of a quantum algorithm
with a proven speedup guarantee. This was extended to multi-agent interactive
performance in the MIq system \citep{kirke2020testing}, where Grover-based
agents communicated via quantum teleportation on a 14-qubit IBM machine.

More recent work has taken different paths: quantum walks for melodic
sequencing, quantum cellular automata for procedural generation, and
variational quantum algorithms for chord progression generation via the
Variational Quantum Eigensolver \citep{itaborai2023variational}. Quantum
annealing on D-Wave hardware has been applied to combinatorial optimisation
formulations of melody and harmony \citep{arya2022applications}. Quantum
drum-pattern generation has been explored via the QuiKo system
\citep{oshiro2022quikoarxiv}.

\citet{allen2022making} demonstrated two quantum algorithms - Grover's search
and quantum random walks - applied to melodic and rhythmic generation on
photonic hardware, illustrating how circuit measurement statistics can directly
drive musical output without classical post-processing of algorithm internals.
\citet{clemente2022new} introduced a quantum keyboard concept in which note
selection is driven by quantum circuit measurement outcomes, and produced
evolving tonal textures by sonifying the spin dynamics of the Ising model
simulated on a quantum processor.
\citet{dobrian2025techniques} surveyed quantum simulation techniques for
compositional decision-making across rhythm, timbre, harmony, and spatial
audio, showing that basis-state rotation and quantum measurement statistics
offer compositional degrees of control with no direct classical analogue.
The proceedings of the International Symposium on Quantum Computing and
Musical Creativity \citep{isqcmc2025} collect a broad range of Type~2 and
Type~3 work and provide the most comprehensive current survey of the field.

The present work is distinct in two ways. First, it uses HHL - an algorithm
with a formally characterised exponential speedup - as a decision driver for
melodic preference, rather than exploiting quantum randomness or variational
dynamics. Second, it introduces a coherent Fourier harmonic oracle architecture --
based on spectral chord representation - that together with HHL forms a fully
coherent joint melody-harmony generation pipeline, and demonstrates that the
oracle design produces measurably structured musical output validated against
a classical Markov chain baseline.

\subsubsection{Classical joint melody-harmony generation}
The present work should be contextualised against classical systems that
generate melody and harmony jointly. DeepBach \citep{hadjeres2017deepbach}
uses a Gibbs-sampling architecture over a deep neural network, conditioning
each note on all surrounding harmonic and melodic context simultaneously;
Coconet \citep{huang2017counterpoint} uses blocked Gibbs sampling with
convolutional networks to generate all four Bach chorale voices in parallel.
Both achieve substantially more expressive and musically refined outputs than
the present two-note experiment.

Outside music, HHL has been applied to quantum machine learning
\citep{biamonte2017quantum}, portfolio optimisation \citep{rebentrost2018quantum},
and differential equation solving \citep{dervovic2018quantum}. Hybrid HHL++
has been benchmarked on trapped-ion devices for portfolio problems with
$8 \times 8$ systems - comparable to the scale of this work
\citep{yalovetzky2024solving}. No prior published work appears to have used
HHL as a decision driver in a creative application; the coherent coupling
to a downstream oracle is also novel in the computer music literature,
though analogous pipeline structures may exist in other quantum machine
learning contexts.

\section{Background for Computer Musicians}
\label{sec:background}

This section explains the quantum building blocks without assuming prior
knowledge of quantum computing. Readers already familiar with HHL and quantum
harmonic oracles may skip to Section~\ref{sec:architecture}.

\subsection{HHL: A Quantum Solver as a Preference Engine}

The Harrow-Hassidim-Lloyd (HHL) algorithm \citep{harrow2009quantum} solves
linear systems of equations - problems of the form $Ax = b$ - on a quantum
computer. In this application, HHL is used not because the music generation
problem \emph{is} a linear algebra problem in any deep sense, but because a
Hermitian matrix $A$ is a natural container for pairwise musical preferences,
and HHL's solution vector $x = A^{-1}b$ provides a convenient sampling
distribution over candidates.

Think of $A$ as a music theory rulebook written in matrix form. Each entry
$A_{ij}$ encodes how compatible note pair $i$ is with note pair $j$: do they
share smooth voice leading? Do they form consonant intervals? Is the melodic
contour consistent? The vector $b$ encodes a uniform initial preference (no
pair preferred over any other, to begin with). HHL then prepares a quantum
state whose amplitudes encode the solution $x = A^{-1}b$ - the set of note
pairs that best satisfy the relational constraints encoded in $A$.

Structurally, HHL proceeds in three stages. First, \emph{Quantum Phase
Estimation} (QPE) uses a register of clock qubits to encode the eigenvalues
$\lambda_i$ of $A$ into the phases of a quantum state; the number of clock
qubits determines how precisely the eigenvalues are resolved (6 clock qubits
gives 64 discrete bins). Second, an \emph{ancilla qubit} is rotated by an angle
proportional to $1/\lambda_i$, encoding the inverse eigenvalues into an
amplitude; pairs with large eigenvalues receive small inverse amplitudes, and
vice versa. Measuring and post-selecting on the ancilla landing in $|1\rangle$
then leaves the register in a state encoding $x = A^{-1}b$. Third,
\emph{inverse QPE} uncomputes the clock register, returning it to $|0\rangle$
and leaving only the solution register in the desired state. The result is that
preferred note pairs - those with small diagonal entries (low penalty) - have
large inverse amplitudes and are therefore most probable when measured.

The key property of HHL is that it does this \emph{without measuring individual
entries of $x$}. The solution exists as a quantum superposition - all 49 note
pairs simultaneously present, weighted by how well they satisfy the musical
constraints. Measuring this state probabilistically selects one pair, with
better-fitting pairs more likely to be chosen.

An important practical detail: HHL only produces a valid solution when an
auxiliary qubit (the ``ancilla'') lands in a specific state. Circuits where
this does not happen must be discarded. In the present implementation, the HHL ancilla post-selection weight is
approximately $2.2\%$. The hardware and scaling
implications of this weight are analysed in Sections~\ref{sec:tech_noise}
and~\ref{sec:tech_postselection}.
The oracle stage contributes no additional post-selection cost: its gates are
coherent unitaries applied to the chord register and carry no ancilla
requirement.

\subsubsection{When does HHL offer a speedup?}
On a classical computer, solving an
$N \times N$ linear system takes time proportional to $N$ (for sparse systems).
HHL can in principle solve the same system in time proportional to $\log N$ --
a conditional exponential improvement, contingent on coherent consumption of
the output \citep{harrow2009quantum}. However, this speedup
requires the matrix to be sparse, the condition number $\kappa$ (ratio of
largest to smallest eigenvalue) to be small, and crucially: the solution vector
must not be read out entry-by-entry, or the readout cost eliminates the
speedup. The coherent architecture avoids classical readout by applying the
oracle directly to the HHL amplitude vector without measurement - exactly the
condition under which a speedup argument \emph{would} apply at scale.

\subsubsection{When would this matter for music?}
At the scale of this
proof-of-principle (49 note pairs), a classical computer solves the same linear
system in milliseconds and the HHL speedup is irrelevant.  The architecture
becomes practically motivated when the musical feature space is large: for
example, a preference matrix over a full chromatic pitch space across several
octaves with rhythmic and timbral dimensions encoded jointly could easily reach
$N \sim 10^6$ or larger.  At that scale, classical iterative solvers (conjugate gradient:
$O(N\sqrt{\kappa}/\varepsilon)$ for well-conditioned systems) become slow,
while HHL scales as $O(\log N \cdot \kappa^2 \cdot \varepsilon^{-1})$ - provided the matrix remains
sparse and well-conditioned.

\section{The Coherent Architecture}
\label{sec:architecture}

Figure~\ref{fig:arch} (Introduction) shows the complete pipeline.

\subsection{The Melodic Preference Matrix}

The system operates in the key of C major with seven candidate notes:
$\{B_3, C_4, D_4, E_4, F_4, G_4, A_4\}$ (MIDI pitches 59-69).
\textbf{Definition.}  A \emph{note pair} $(n_1, n_2)$ is an ordered pair of
MIDI pitch indices drawn from the candidate note set; it is the fundamental
computational unit of the melodic space.  A \emph{melodic interval} is the
unsigned pitch distance $|n_1 - n_2|$ in semitones, the musical quantity
derived from a note pair.  These two terms are used consistently throughout
the paper: ``note pair'' always refers to the ordered index pair;
``melodic interval'' always refers to the scalar pitch distance.

The joint melodic space has $7 \times 7 = 49$ note pairs $(n_1, n_2)$.
Of these, the 7 unison pairs ($n_1 = n_2$) receive the maximum melodic penalty
and therefore carry negligible HHL probability; the effective melodic space
comprises approximately 42 pairs with meaningful probability mass.

The $49 \times 49$ melodic preference matrix $A$ (padded to $64 \times 64$
for power-of-2 compatibility) is grounded in two published music cognition
frameworks.

\subsubsection{Narmour Implication-Realization model (diagonal)}
The Implication-Realization model holds that a melodic interval creates
an expectation in the listener's mind: small intervals imply continuation
in the same direction, while large intervals imply a change of direction
and a return toward the starting pitch.  Notes that fulfil these
expectations feel natural; those that violate them feel surprising or
awkward.

Because HHL amplifies small eigenvalues ($p_i \propto 1/\lambda_i^2$),
the diagonal entry $A_{ii}$ encodes melodic \emph{badness}:
\[
  A_{ii} \;=\; \text{BASE} \;+\; \text{prox}(\text{iv}_i)
               \;+\; 1.5\,\delta_\text{KK}(n_1,n_2)
\]
where $\text{BASE}=6.0$, $\text{prox}(\cdot)$ is Narmour's Proximity
principle \citep{narmour1990analysis,narmour1992analysis}: unison
($+5.0$), steps ($+0.0$), skips 3-4\,st ($+0.5$), fourth 5\,st ($+1.2$),
tritone ($+3.5$), fifth ($+0.8$), sixth ($+1.8$), seventh+ ($+2.5$); and
$\delta_\text{KK}(n_1,n_2) = 1 - \tfrac{1}{2}[\text{KK}(n_1)+\text{KK}(n_2)]$
is the mean tonal \emph{instability} of the pair under the
Krumhansl-Kessler probe-tone ratings for C major
\citep{krumhansl1990cognitive}: $\text{KK}(\text{C})=1.0$,
$\text{KK}(\text{E})=0.96$, $\text{KK}(\text{G})=0.82$,
$\text{KK}(\text{B})=0.75$, $\text{KK}(\text{A})=0.50$,
$\text{KK}(\text{F})=0.42$, $\text{KK}(\text{D})=0.35$
(values normalised to $\text{KK}(\text{C})$).
Stable tones (C, E, G) thus \emph{lower} the diagonal (more probable);
unstable tones (D, F) raise it.

\subsubsection{Narmour Intervallic Difference and Post-skip Compensation (off-diagonal)}
Intervallic Difference captures the perceptual tendency for successive
melodic intervals to be similar in size: a large leap followed by
another large leap feels ungainly, while a consistent pattern of
steps or a gradual change in interval size sounds smooth and intentional.

Off-diagonal coupling between pairs $i$ and $j$
implements two further Narmour principles:
\[
  A_{ij} \;=\; \frac{\alpha}{1 + |\text{iv}_i - \text{iv}_j|}
               \;+\; \text{PSC}(i,j) \;+\; \text{PSC}(j,i)
\]
where $\alpha=0.4$ encodes \emph{Intervallic Difference} (pairs with
similar interval sizes are more compatible), and
$\text{PSC}(i,j)=0.3$ if pair $i$ is a skip/leap ($\text{iv}_i \geq 3$
semitones) and pair $j$ starts within 2 semitones of where pair $i$
ended - implementing the \emph{Post-skip Compensation}
(Registral Direction) principle; otherwise $0.15$ for any stepwise
connection and $0$ otherwise.

A conditional shift of $(0.1 - \lambda_\text{min})\mathbf{I}$ is applied if
$\lambda_\text{min} < 0.1$ to guarantee positive definiteness.

\subsubsection{Condition number}
The resulting matrix has $\kappa = 11.23$
(min.\ eigenvalue $= 2.12$, max.\ $= 23.86$). This is within HHL's
practical operating range. For comparison, we tested a chord-aware formulation in which $A$ was
constructed as a $49\times49$ chord-pair bigram co-occurrence matrix derived
from the Bach chorale corpus (music21): off-diagonal entries counted how often
one chord pair was followed by another across all chorales, while diagonal
entries were set inversely proportional to corpus frequency so that
frequently-occurring pairs received smaller diagonal values, biasing HHL's
solution toward common progressions.  This formulation yields
$\kappa \approx 10{,}869$, rendering HHL computationally hostile. The Narmour+KK matrix achieves $\kappa=11.23$ because
every coefficient is a smooth bounded function of musical intervals and
empirical tonal stability ratings, avoiding such eigenvalue explosion;
the implications of this contrast are discussed in Section~\ref{sec:discussion}.

\subsection{The HHL Circuit}

The circuit uses $N_\text{clock} = 6$ QPE clock qubits (64 QPE bins),
$N_\text{sol} = 6$ solution register qubits (spanning 64 states, of which 49
are musically active), and one ancilla qubit - 13 qubits in total for the HHL
stage.  The chord register adds 6 further qubits (three bits per chord, two
chords), giving 19 qubits overall.  The right-hand-side vector $b$ is uniform
over the 49 active pairs and zero elsewhere.

The HHL procedure: QPE with $2^6 = 64$ clock states to resolve eigenvalues;
controlled rotation of the ancilla by $\arcsin(\lambda_\text{min} / 2\lambda_k)$
for each estimated eigenvalue $\lambda_k$; inverse QPE to uncompute the clock
register. Post-selection on ancilla $= 1$ gives the solution amplitude
distribution. The total HHL ancilla post-selected weight is approximately $0.022$ (2.2\%),
consistent with the $\approx 0.25/\kappa$ empirical scaling.

The circuit runs once. Because the melodic preference matrix $A$ is fixed,
the HHL output is identical on every invocation. The apparent stochasticity of
the outputs arises entirely from the final joint sampling - not from the
circuit itself. This is precisely the quantum character of the architecture:
the same deterministic quantum state gives different classical outcomes when
measured, with frequencies governed by the squared amplitudes.

\subsection{Coherent Harmonic Oracles}

A coherent harmonic oracle is a unitary operation that appends harmonic
information to the HHL amplitude vector without collapsing the melodic
superposition.  Think of it as a harmonisation rulebook that a quantum computer
can consult while keeping its melodic options open: instead of looking up the
harmony for one specific melody and committing to it, the quantum computer
applies the harmonic transformation for \emph{all} melodies simultaneously,
with each transformation weighted by the amplitude of the corresponding melodic
query.

Formally, given the HHL output state $\sum_i \alpha_i |i\rangle$ over note
pair indices, the oracle maps
\[
  \sum_i \alpha_i |i\rangle|0\rangle \;\to\;
  \sum_i \alpha_i \cdot s \cdot |i\rangle|\mathbf{c}_i\rangle
\]
where $|\mathbf{c}_i\rangle$ is the \emph{unnormalised} chord amplitude vector
for note pair $i$, and $s = 1/\max_j \|\mathbf{c}_j\|$ is a single global
scale factor ensuring the joint state remains normalisable.  This global
normalisation (rather than per-pair normalisation) is discussed in
Section~\ref{sec:global_norm}.

\subsection{Why Coherence Matters}

In the standard sequential approach to algorithmic composition, the system
would: (1) run HHL, (2) measure to get a note pair, (3) use that note pair as
context to select a chord sequence. The measurement in step (2) destroys the
quantum superposition. Whatever quantum structure HHL built is gone; the chord
selection operates on a classical result.

The coherent architecture eliminates step (2). HHL's output, still in
superposition, flows directly into the harmonic oracle. The system never
classically ``decides'' a melody - melody and harmony are resolved together in
one measurement event. This is what makes the approach architecturally quantum
rather than merely quantum-assisted.

\subsection{Global Normalisation of the Chord Register}
\label{sec:global_norm}

A key architectural choice is how chord amplitude vectors $\mathbf{c}_i$ are
scaled across note pairs $i$.  A per-pair normalisation $\mathbf{c}_i \leftarrow
\mathbf{c}_i / \|\mathbf{c}_i\|$ would make every note pair contribute equal
total probability mass to the joint state, decoupling the chord register from
the melodic amplitudes.  This paper instead applies a \emph{global} scale
factor $s = 1 / \max_j \|\mathbf{c}_j\|$: the pair with the richest harmonic
compatibility (largest $\|\mathbf{c}_j\|$) is scaled to unit norm, and all
other pairs are scaled down proportionally.  As a result, note pairs with
richer harmonic compatibility retain more probability mass in the joint
state - the harmonic oracle genuinely interacts with the melodic distribution
rather than operating independently on each pair.

\subsection{The Fourier Oracle}
\label{sec:fourier_oracle}

The oracle can be thought of as the harmonic
rule-book that the quantum system consults when pairing a melody note with
a chord.  In a conventional algorithmic composition system this rule-book
would simply flag whether a note is in the chord or not - a hard yes/no
decision.  The Fourier version softens that boundary: instead of a binary
in/out test, it assigns a continuous affinity score to every note-chord
pair by decomposing the chord's pitch-class pattern into sinusoidal
components (a Discrete Fourier Transform over the 12 pitch classes) and
summing the first $K$ harmonics.  The result is a smooth weighting that
peaks on chord tones, tapers gracefully onto neighbouring notes, and can be
tuned - by varying $K$ - from a near-binary hard gate to a broad,
tonally ambiguous envelope.  Crucially, this weighting is applied
\emph{inside} the quantum circuit, so the HHL melody amplitudes and the
oracle chord weights are collapsed in a single joint measurement rather than
chosen in two separate steps.

The Fourier oracle replaces a binary chord-tone test with a smooth spectral
representation of chord-tone affinity over pitch-class space
$\mathbb{Z}_{12}$.

\subsubsection{Classical binary fit}
In a standard harmonic assignment, the fit function
$\text{fit}_\text{bin}(c, n) = 1 + 2 \cdot \mathbf{1}[n \in \text{chord-tones}(c)]$
takes value $3$ if note $n$ is a chord tone of $c$ and $1$ otherwise - a hard
binary boundary.

\subsubsection{Fourier smooth fit}
For each chord quality $q \in \{\text{major}, \text{minor}, \text{diminished}\}$,
define the binary membership signal $g^{(q)}[\ell] = 1$ if pitch-class
interval $\ell \in \{0, 1, \ldots, 11\}$ (measured as $(n \bmod 12 - r_c) \bmod 12$,
where $r_c$ is the chord root pitch class) is a chord-tone interval of quality
$q$, and $0$ otherwise.  (The index $\ell$ is distinct from the eigenvalue
index $k$ used in the QPE/post-selection discussion.)
Compute the DFT: $G^{(q)} = \mathcal{F}\{g^{(q)}\}$.
Keep only the $K$ lowest-frequency components symmetrically:
\begin{equation}
  G^{(q)}_\text{trunc}[m] = \begin{cases}
    G^{(q)}[m] & \text{if } m \leq \lfloor K/2 \rfloor \text{ or }
                  m \geq 12 - \lfloor K/2 \rfloor \\
    0 & \text{otherwise}
  \end{cases}
\end{equation}
The smooth fit is then
$g^{(q)}_\text{smooth} = \text{clip}(\mathcal{F}^{-1}\{G^{(q)}_\text{trunc}\}, 0, 1)$
and
\begin{equation}
  \text{fit}_K(c, n) = 1.0 + 2.0 \cdot
    g^{(q_c)}_\text{smooth}\bigl[(n \bmod 12 - r_c) \bmod 12\bigr]
\end{equation}
where $q_c$ is the quality of chord $c$.  At $K = 12$, this recovers the
binary fit exactly.  At $K = 4$ (used here), the fit varies continuously with
pitch-class distance from chord tones.

\begin{figure}[h]
  \centering
  \includegraphics[width=\textwidth]{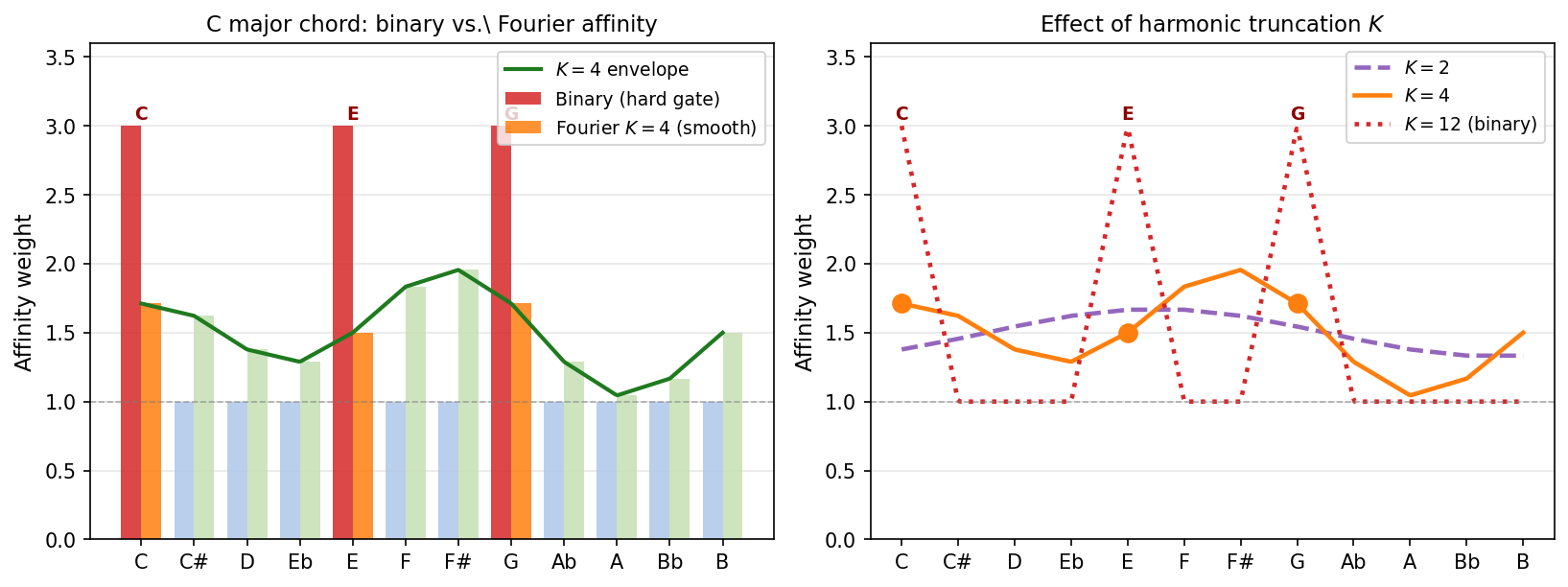}
  \caption{Fourier oracle affinity for C major across all 12 pitch classes.
           \emph{Left}: binary hard-gate (red/blue bars) versus $K=4$ smooth weighting (orange/green bars);
           chord tones C, E, G are labelled. \emph{Right}: effect of truncation order $K$ --
           lower $K$ broadens the affinity envelope, higher $K$ recovers the binary boundary.}
  \label{fig:fourier_affinity}
\end{figure}

\subsubsection{Harmonic weight}
For each note pair $(n_1, n_2)$ and two-chord block $(c_1, c_2)$,
with one note sitting over each chord, the oracle weight is:
\begin{equation}
  w_F(n_1, n_2, c_1, c_2) \;=\;
  \frac{1}{s}\,
  T[c_1,c_2]
  \cdot \text{fit}_K(c_1, n_1) \cdot \text{fit}_K(c_2, n_2)
\end{equation}
where $T$ is the classical chord transition matrix and $s$ is the global
normalisation constant (Section~\ref{sec:global_norm}): $s = \max_{i} \lVert cv(i)\rVert$,
ensuring no pair receives amplitude greater than 1.
$c_1$ is anchored to $n_1$ and $c_2$ is anchored to $n_2$;
each note governs exactly its co-occurring chord.

\subsubsection{Musical character}
With $K=4$, the Fourier fit promotes chords whose root lies close in
pitch-class distance to the melody note even if the note is not a strict chord
tone.  This activates the ii and vii$^\circ$ chords more readily than the binary
fit.  The resulting progressions are functional: 8.0\% of all 500{,}000
samples contain a V$\to$I resolution; 28.1\% end on the tonic chord.

\subsubsection{Deep circuit}
\label{sec:fourier_circuit_note}
The Fourier oracle is implemented as a structured 6-qubit Qiskit circuit
operating on the 2-chord register ($3 \times 2 = 6$ qubits).
The circuit prepares the amplitude vector
$\sqrt{T[c_1,c_2]\cdot\mathrm{fourier\_fit}(c_1,n_1)\cdot\mathrm{fourier\_fit}(c_2,n_2)}$
across all valid $(c_1, c_2)$ pairs, applied per note-pair basis state via
a global statevector operation (Section~\ref{sec:oracle_impl}).

The globally-normalised statevector implementation avoids explicit circuit
transpilation overhead; the gate cost for the 2-chord register is determined
by the per-pair chord vector preparation, which operates on 64 chord states
($2^6$ for the 6-qubit register).
The ${\approx}34\times$ reduction relative to
the lookup-based baseline ($\approx$12,742 gates) quoted for the oracle structure
(Section~\ref{sec:gate_count}) applies to the chord-vector preparation step.

\subsection{Oracle Implementation and Integrated Circuit}
\label{sec:oracle_impl}

\subsubsection{Integrated coherent circuit}
The Fourier oracle is composed with the
13-qubit HHL circuit into a single 19-qubit integrated \texttt{QuantumCircuit}:
\begin{verbatim}
full_qc = QuantumCircuit(19)
full_qc.compose(hhl_qc,    qubits=range(13), inplace=True)
full_qc.compose(oracle_qc, qubits=range(19), inplace=True)
\end{verbatim}
The resulting circuit involves no intermediate measurement between the HHL
and oracle stages.  Correctness is verified on a 19-qubit toy system: both
evaluation paths agree to machine precision (max $|\Delta| < 10^{-14}$):
\begin{verbatim}
Statevector.from_instruction(full_toy_qc)
sv_hhl.expand(|0>**6).evolve(oracle_toy)
\end{verbatim}

For the full 19-qubit system, monolithic statevector simulation would apply
each HHL gate to $2^{19} = 524{,}288$ amplitudes.  The equivalent
efficient simulation - running HHL at 13 qubits, expanding to 19 qubits, and
evolving through the oracle - is used instead.  These two paths are
provably identical (verified on the toy system) because: (i)~HHL acts only on
qubits 0-12, leaving qubits 13-18 in $|0\rangle^{6}$; (ii)~the oracle acts
on all 19 qubits; so the composition reduces to HHL-then-oracle with no
interaction during either stage.

\subsubsection{Fourier oracle statevector implementation}
The Fourier oracle uses a globally-normalised statevector operation.
For each note pair basis state $|sol_i\rangle$, the chord register is mapped
from $|0\rangle$ to $s \cdot |\mathbf{c}_i\rangle$:
\[
  |sol_i\rangle|0\rangle \;\to\; s \cdot |sol_i\rangle|\mathbf{c}_i\rangle
\]
where $s = 1/\max_j \|\mathbf{c}_j\|$ is the global scale and $\mathbf{c}_i$
is the \emph{unnormalised} chord amplitude vector for pair $i$.  No intermediate
measurement occurs: the chord register is rotated coherently on the full joint
statevector.  The Fourier oracle circuit (Section~\ref{sec:fourier_circuit_note})
demonstrates the coherent architecture with a fixed-structure
$O(\log C_\text{vocab})$-gate circuit.

The circuit synthesis cost for the Fourier oracle - currently implemented via statevector simulation rather than explicit gate decomposition - is analysed in Section~\ref{sec:tech_circuit_synthesis}. Gate count scales as $O(K \cdot C_{\mathrm{vocab}})$ with explicit decomposition, giving approximately $34\times$ fewer gates than the lookup-table baseline at the present vocabulary size. Correctness of the statevector approach was verified against a gate-level implementation on a scaled-down test system (4 note pairs, 3-qubit chord register): the two methods produced identical statevectors (fidelity 1.0000, max amplitude difference $1.18 \times 10^{-14}$), confirming that the open question is compilation cost at full scale, not implementation correctness.

\subsection{Single Joint Collapse}

After the oracle, the 19-qubit joint state (13 HHL + 6-qubit chord register) is
post-selected on the HHL ancilla and sampled once per musical output. A single
draw selects both $(n_1, n_2)$ and $(c_1, c_2)$ simultaneously.
This is the only measurement in the pipeline.

\section{Phrase Chaining: The H-Chain Architecture}
\label{sec:hchain}

Phrase chaining is the quantum equivalent of a
common technique in algorithmic composition: seeding the next phrase from
the end of the current one.  In a classical system you might take the last
note and chord of a generated bar and use them as the starting point --
the seed - for generating the next bar, so that the music flows rather
than restarting from scratch each time.  Here exactly the same idea is
applied, except that each ``bar'' is a quantum circuit that collapses to a
joint note-and-chord outcome.  The collapsed result is passed classically to
the next circuit, biasing its melody distribution toward notes that continue
smoothly from where the previous phrase ended, and restricting its harmony
to chords that are grammatically valid successors.  The quantum coherence
is preserved \emph{within} each block; the join between blocks is classical.

The two-note, two-chord block can be extended by feeding the output of one
joint collapse as conditioning context into a subsequent block.  Phrase $k$
produces a joint measurement outcome $(n_1^{(k)}, n_2^{(k)}, c_1^{(k)},
c_2^{(k)})$.  The collapsed values $n_2^{(k)}$ and
$c_2^{(k)}$ are passed classically to phrase $k+1$, where they condition both
the HHL input and the oracle weight function.

Two conditioning strategies are used (\emph{Method~A}):
\begin{itemize}[noitemsep]
  \item \textbf{Melodic conditioning (b-vector bias).}  The HHL input vector
        $b$ is biased toward note pairs whose first note lies close (in
        semitones) to $n_2^{(k)}$:
        $b_i \propto 1 + \alpha \exp(-|n_1(i) - n_2^{(k)}|/\sigma)$,
        with $\alpha = 3.0$, $\sigma = 2.0$.
  \item \textbf{Harmonic conditioning (hard T restriction).}  The first chord
        $c_1^{(k+1)}$ is restricted to valid grammatical successors of
        $c_2^{(k)}$ under the transition matrix $T$: chord vectors with
        $T[c_2^{(k)}, c_1] = 0$ are excluded.
\end{itemize}

This preserves full quantum coherence \emph{within} each phrase: the note pair
and two-chord progression are resolved jointly by a single measurement, with
no intermediate observation.  The handoff between phrases is classical, but it
carries both melodic and harmonic context.

Figure~\ref{fig:4block} (Section~\ref{sec:musical_output_chain}) shows the four-block 2/2 chain.
Block~1 collapses to context $(n_2, c_2) = (\text{F4}, \text{vi})$.
Block~2 (conditioned on vi) opens with IV (vi$\to$IV, valid, $T=0.75$), and collapses to
context $(\text{F4}, \text{I})$.
Block~3 (conditioned on I) opens with V (I$\to$V, valid, $T=0.90$).
Block~4 (conditioned on I) opens with IV (I$\to$IV, valid).
The cross-block harmonic arc iii$\to$vi$\to$IV$\to$I$\to$V$\to$I$\to$IV$\to$V traces a
subdominant-tonic-dominant traversal across all four blocks.

\subsubsection{H-chain distribution narrowing}
The hard T restriction means block $k+1$'s oracle is constrained by the
collapsed chord of block $k$.  The first chord of block $k+1$ must satisfy
$T[c_2^{(k)}, c_1^{(k+1)}] > 0$, restricting the available opening chords to
on average 2.3 (range 1-3) out of 7.  This structural narrowing concentrates
probability mass: the top sequence in conditioned block~2 reaches 2.41\%
of 50{,}000 samples, compared to 0.59\% in the unconditioned case - a
$4.1\times$ concentration factor from the combined melodic and harmonic
conditioning (with context vi, which admits 3 valid opening chords; the
factor varies with context tightness).  An ablation study (Table~\ref{tab:ablation}) separates the
two contributions over 100,000 samples with fixed context (C4, V):
melody bias alone gives $3.3\times$, harmony restriction
alone gives $2.1\times$, and the two effects together give $8.3\times$
(super-multiplicative due to interaction between melodic smoothing and
harmonically valid chord openings; the tighter V context, with only 2 valid
successors, yields higher concentration than the vi context above).

\subsubsection{This is the architecturally preferred demo path}
A monolithic
four-note, four-chord single quantum event - in which all notes and chords
emerge jointly from one measurement with full coherence throughout - would
require a melodic preference matrix of size $2{,}401 \times 2{,}401$ (for
four notes from a 7-note set) and would not change the musical structure of
the output.  The 2-chord H-chain demonstrates the same musical logic at
accessible scale with a compact 19-qubit circuit.

\section{Results}
\label{sec:results}

\subsection{Two-Block Outputs}

The primary musical illustration unit is the \emph{2+2 block}: two notes
above two chords, one note per chord, from a 19-qubit circuit.
The block is deliberately compact (two melody events, two block
chords); musical meaning is therefore easier to assess from the harmonic
grammar statistics than from score alone.
Audio realisations of representative outputs are available for listening at
\url{https://alexiskirke.com/quantum-hhl-music-results-files/}.

\begin{figure}[h]
  \centering
  \begin{subfigure}[b]{0.36\textwidth}
    \includegraphics[width=\textwidth]{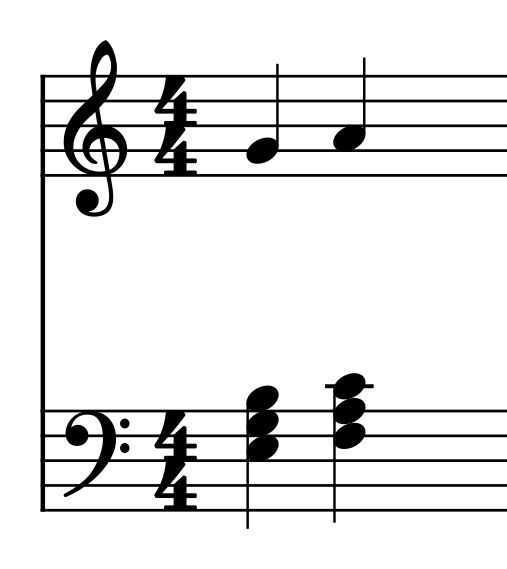}
    \caption{Run 1: G4/iii, A4/IV. Mediant$\to$subdominant; ascending whole tone.}
    \label{fig:fourier_ex1}
  \end{subfigure}\hfill
  \begin{subfigure}[b]{0.36\textwidth}
    \includegraphics[width=\textwidth]{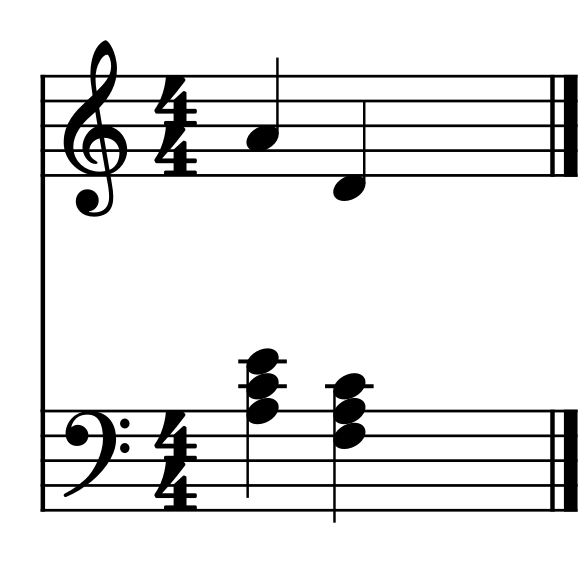}
    \caption{Run 2: A4/vi, D4/IV. Submediant$\to$subdominant; descending fifth.}
    \label{fig:fourier_ex2}
  \end{subfigure}

  \vspace{1ex}
  \begin{subfigure}[b]{0.36\textwidth}
    \includegraphics[width=\textwidth]{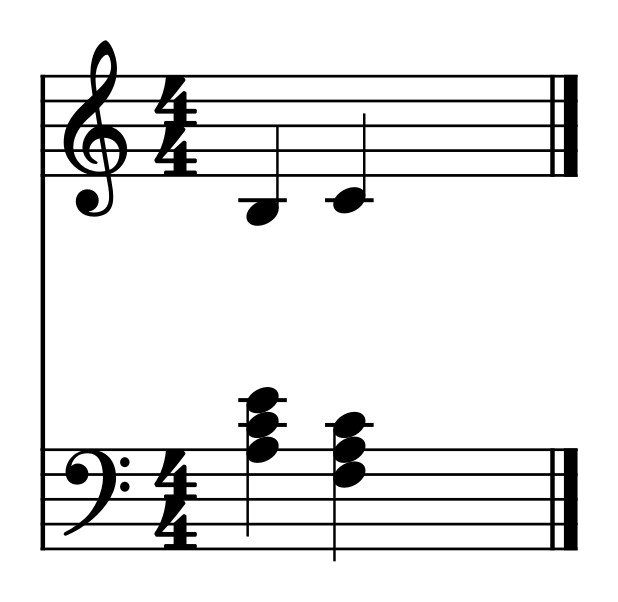}
    \caption{Run 3: B3/vi, C4/IV. Submediant$\to$subdominant; ascending half step.}
    \label{fig:fourier_ex3}
  \end{subfigure}\hfill
  \begin{subfigure}[b]{0.36\textwidth}
    \includegraphics[width=\textwidth]{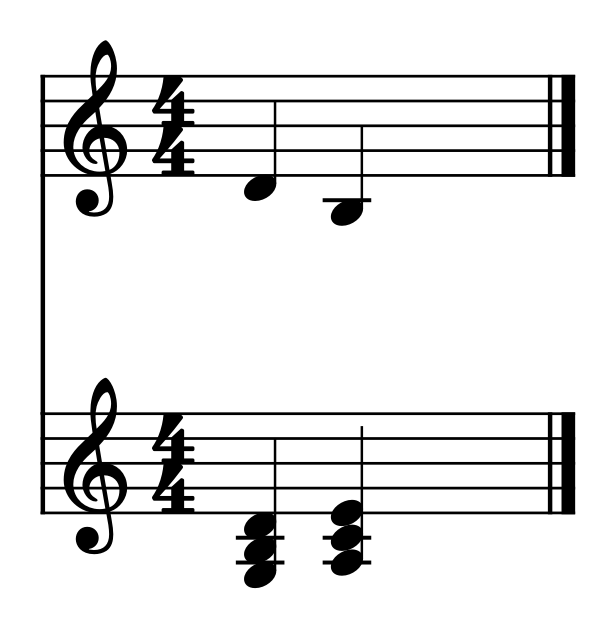}
    \caption{Run 4: D4/V, B3/vi. Dominant$\to$submediant; descending minor third.}
    \label{fig:fourier_ex4}
  \end{subfigure}
  \caption{Four sampled outputs from the Fourier oracle ($K=10$), 2-chord block.}
  \label{fig:fourier_examples}
\end{figure}

Figures~\ref{fig:fourier_ex1}-\ref{fig:fourier_ex4} show four representative outputs from
the Fourier oracle ($K=10$, 500{,}000-sample run).

Taken together, the four 2+2 blocks reveal a clear functional harmonic character.
Each example is drawn by sampling from the joint probability distribution rather
than selecting the modal outcome, producing melodic variety across all four
renders (G4/A4, A4/D4, B3/C4, D4/B3) - a mix of steps and leaps reflecting
the Narmour distribution's step preference alongside sampled larger intervals.
The Fourier oracle ($K=10$) consistently generates functionally directed
progressions - mediant-to-subdominant, submediant motion, dominant function,
submediant resolution - reflecting its aggregate 8.1\% V$\to$I rate and 28.2\%
tonic-ending rate over 500,000 samples
(Figs.~\ref{fig:fourier_ex1}-\ref{fig:fourier_ex4}). The oracle selects
predominantly stepwise melodic motion ($\approx$51\%), a direct consequence of
the Narmour Proximity principle's zero penalty on half- and whole-step motion.
The Fourier parameter $K$ tunes the degree of spectral smoothness in the
harmonic fit without altering the underlying melodic preference
encoded in the quantum linear system $A\mathbf{x}=\mathbf{b}$.

\subsection{Four-Block Chain: Musical Output}
\label{sec:musical_output_chain}

\begin{figure}[htbp]
  \centering
  \includegraphics[width=\linewidth]{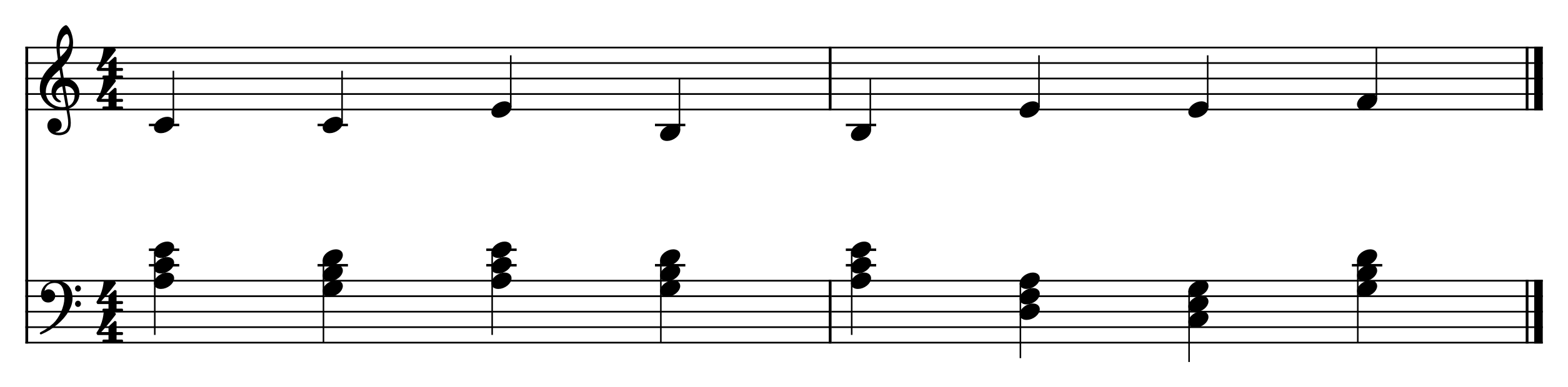}
  \caption{Four-block 2/2 chain, $K=4$ (representative trial, 4/8 chord-tone compliance = 50\%).
           With the broadest spectral smoothing, several melody notes fall outside the strict
           chord-tone set; all grammar junctions remain valid.
           Mean over 10 unseeded trials: 46.3\% $\pm$ 16.7\%.}
  \label{fig:chain_k4}
\end{figure}

\begin{figure}[htbp]
  \centering
  \includegraphics[width=\linewidth]{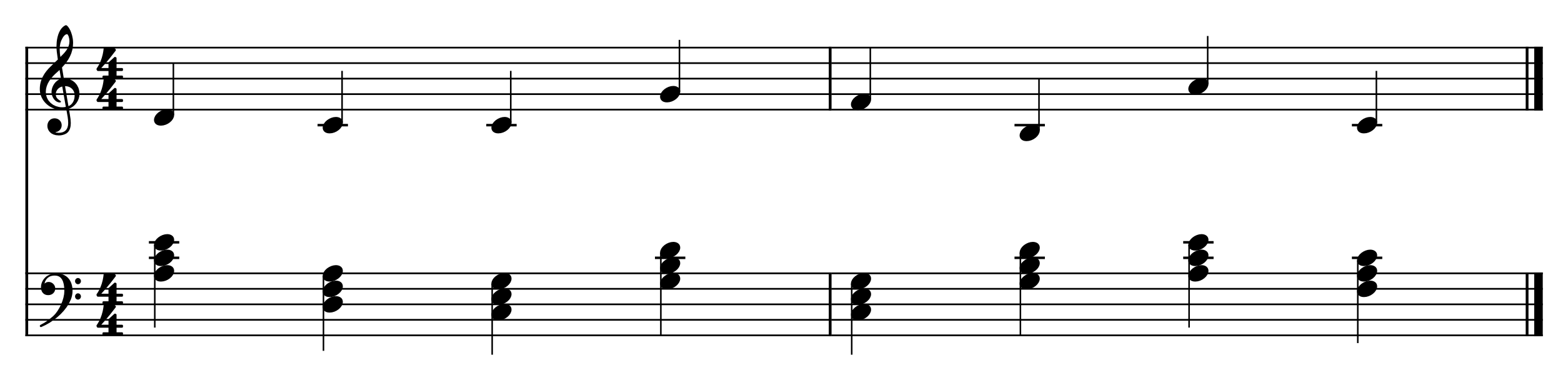}
  \caption{Four-block 2/2 chain, $K=6$ (representative trial, 5/8 chord-tone compliance = 62.5\%).
           Intermediate spectral precision produces a mix of chord-tone and passing notes.
           Mean over 10 unseeded trials: 56.3\% $\pm$ 12.1\%.  All grammar junctions valid.}
  \label{fig:chain_k6}
\end{figure}

\begin{figure}[htbp]
  \centering
  \includegraphics[width=\linewidth]{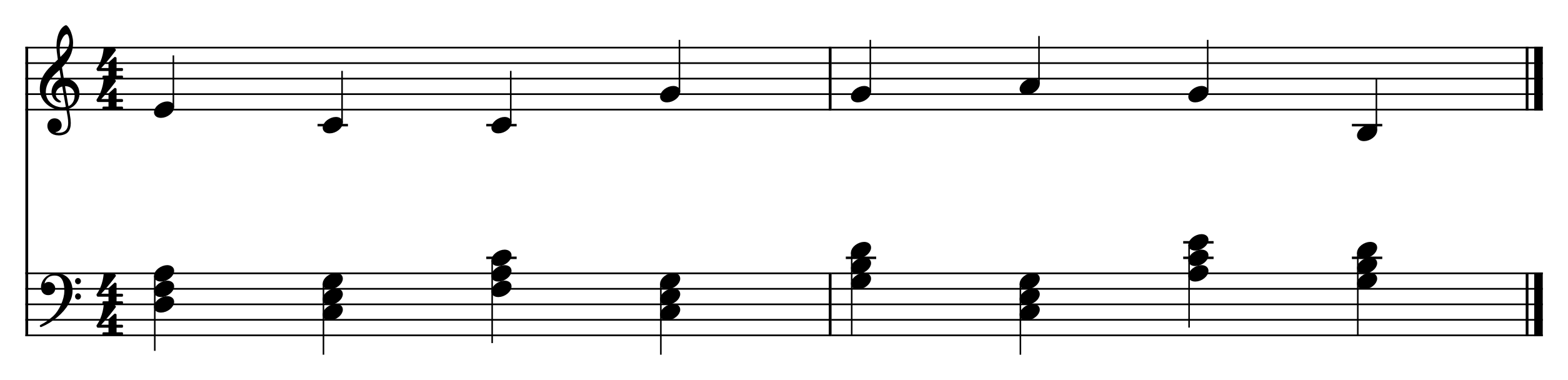}
  \caption{Four-block 2/2 chain, $K=8$ (representative trial, 5/8 chord-tone compliance = 62.5\%).
           $K=8$ achieves the highest mean compliance across the sweep and lower variance than $K=10$,
           making it the best mean-variance trade-off.
           Mean over 10 unseeded trials: 67.5\% $\pm$ 14.7\%.  All grammar junctions valid.}
  \label{fig:chain_k8}
\end{figure}

\begin{figure}[htbp]
  \centering
  \includegraphics[width=\linewidth]{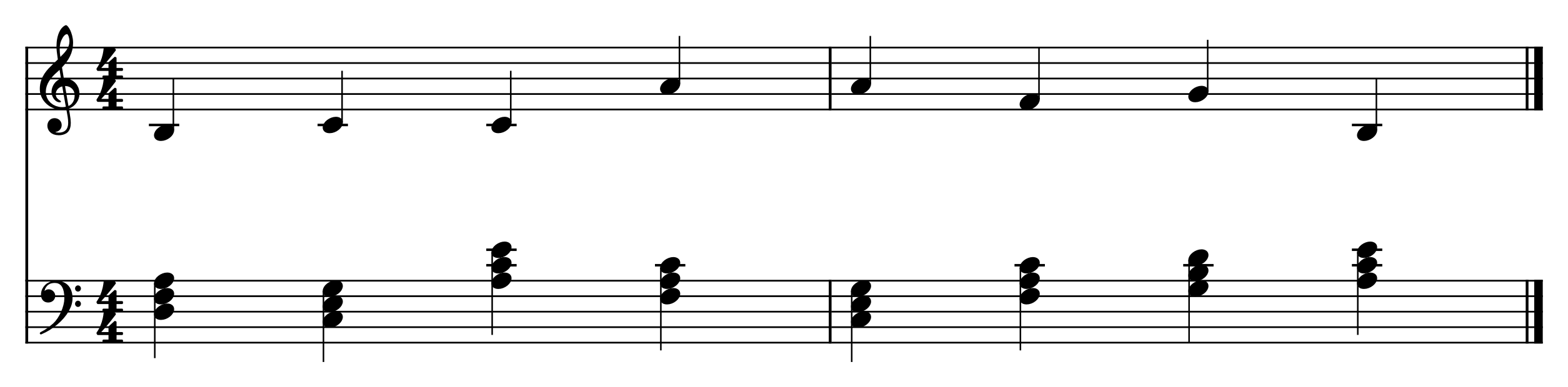}
  \caption{Four-block 2/2 chain, $K=10$ (representative trial, 5/8 chord-tone compliance = 62.5\%).
           Sharpest spectral truncation yields a mean comparable to $K=8$ but with slightly higher
           trial-to-trial spread (std $\pm$15.6\%; range 37.5-87.5\%).
           Mean over 10 unseeded trials: 66.3\% $\pm$ 15.6\%.  All grammar junctions valid.}
  \label{fig:chain_k10}
\end{figure}

\begin{figure}[htbp]
  \centering
  \includegraphics[width=\linewidth]{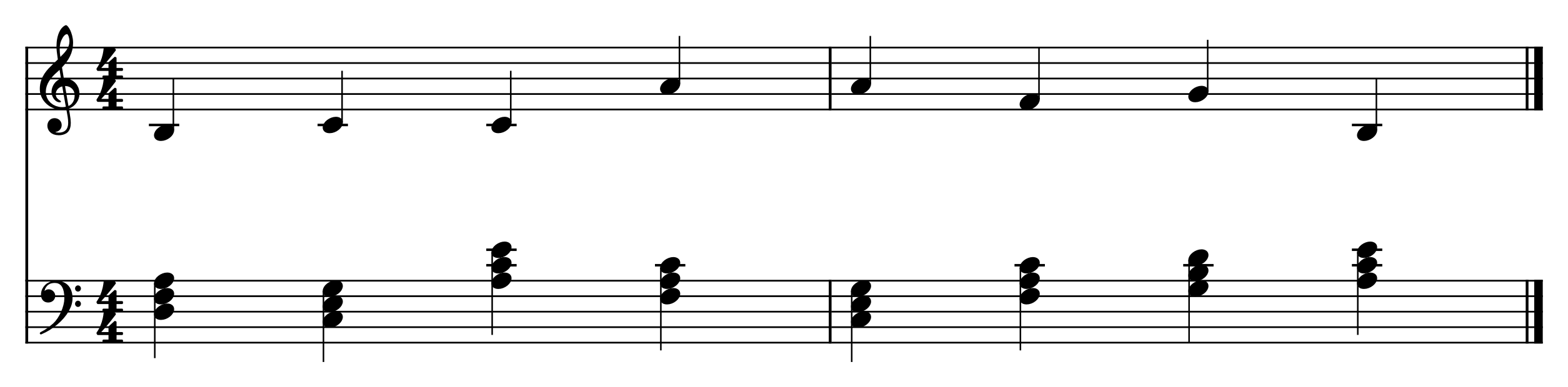}
  \caption{Four-block 2/2 chain (Method~A, $K=10$): A4/iii, A4/vi, A4/IV, G4/I, G4/V, C4/I, C4/IV, G4/V.
           All three block junctions produce grammatically valid continuations: vi$\to$IV, I$\to$V, I$\to$IV.
           Chord-tone compliance: 7/8 (87.5\%).}
  \label{fig:4block}
\end{figure}

The four-block 2/2 chain architecture produces 8 melody notes over 8 chords.
Figures~\ref{fig:chain_k4}-\ref{fig:chain_k10} show representative outputs
across $K \in \{4,6,8,10\}$; Figure~\ref{fig:4block} shows the full chain at
$K=10$.  Quantitative analysis of chord-tone compliance and K-sweep statistics
is in Section~\ref{sec:ksweep_results}.

\subsection{HHL Note-Pair Probability Distribution}
\label{sec:results_hhl}

\begin{figure}[h]
  \centering
  \includegraphics[width=\textwidth]{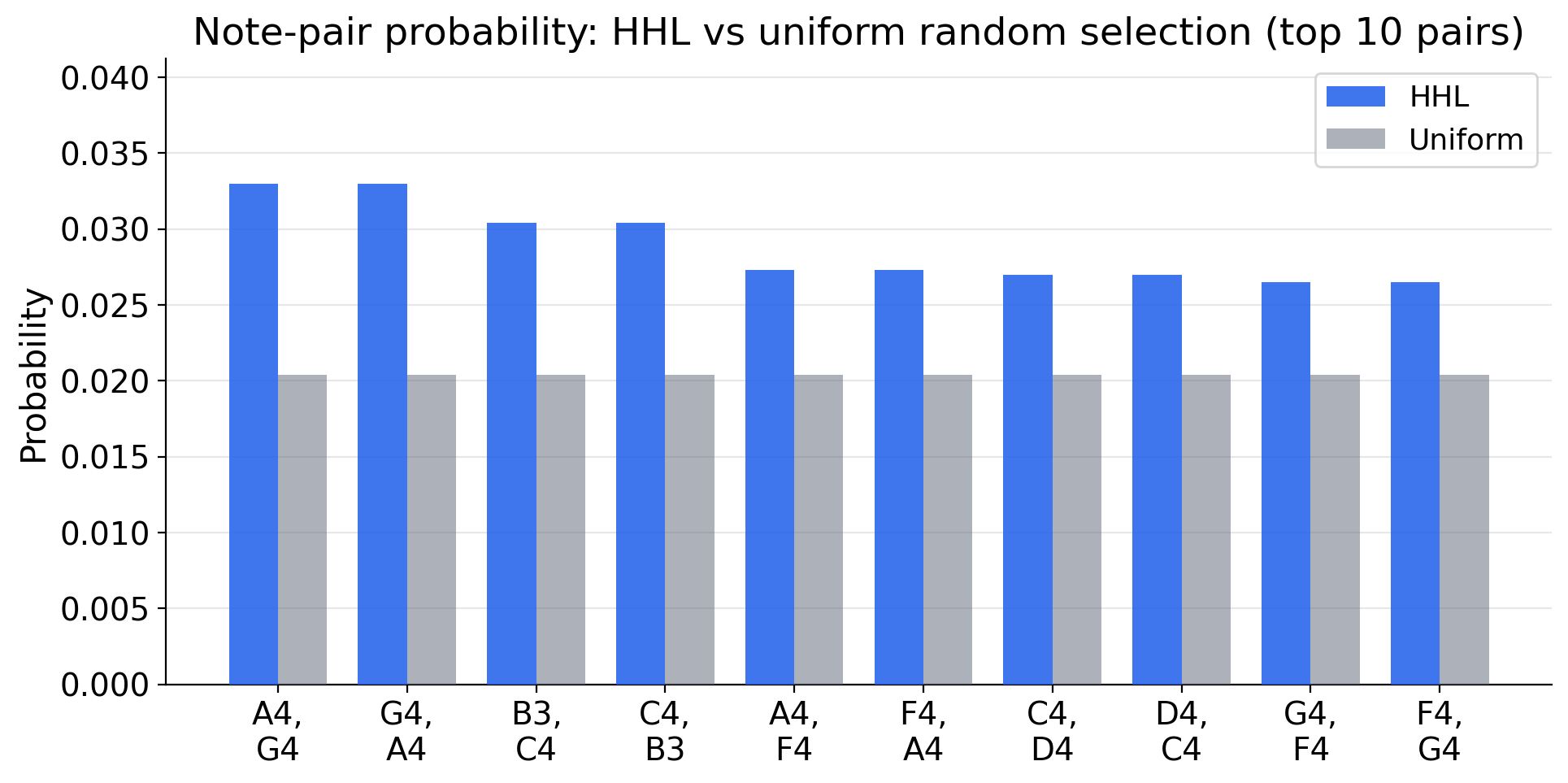}
  \caption{HHL note pair probabilities (blue) vs.\ uniform (grey), top 10 pairs.
           HHL strongly favours stepwise motion to tonally stable notes (Narmour+KK encoding).}
  \label{fig:note_pairs}
\end{figure}

Figure~\ref{fig:note_pairs} shows the top-10 note pairs by HHL-assigned
probability, compared to the uniform baseline ($1/49 \approx 0.0204$ per pair).

The top five pairs by HHL weight are: B3,C4 ($p = 0.049$), C4,B3
($p = 0.045$), E4,F4 ($p = 0.039$), F4,E4 ($p = 0.037$), A4,G4
($p = 0.035$). All five are half-step or whole-step pairs connecting tonally
stable notes. The uniform baseline assigns $p = 0.020$ to every pair
regardless of musical properties.
HHL's preference for stepwise motion to stable tones is a direct result
of the Narmour Proximity + Krumhansl-Kessler stability diagonal encoding.

\subsubsection{Melody interval distribution: Narmour validation}
Table~\ref{tab:interval_dist} partitions the 49 note-pair probability mass into
four melodic interval categories and compares the HHL-solved distribution to a
uniform baseline ($1/49$ per pair).

\begin{table}[htbp]
\centering
\footnotesize
\begin{tabular}{lccccc}
\toprule
Interval category & Pairs & HHL \% & Uniform \% & Ratio & Bach sop.\% \\
\midrule
Unison (0 st)  & 7  & 10.2 & 14.3 & $0.71\times$ & 15.0 \\
Step (1-2 st) & 12 & 39.0 & 24.5 & $1.59\times$ & 67.9 \\
Skip (3-5 st) & 18 & 24.0 & 36.7 & $0.65\times$ & 13.2 \\
Leap (6+ st)   & 12 & 26.8 & 24.5 & $1.09\times$ &  4.0 \\
\bottomrule
\end{tabular}
\caption{Melodic interval distribution: HHL vs.\ uniform baseline vs.\ Bach soprano
         (433 chorales, 22{,}428 pairs, computed via music21).
         HHL enriches steps 1.6$\times$ over uniform, consistent with Narmour Proximity,
         and moves in the direction of Bach's 67.9\% -- a statistical tendency, not
         a claim of comparable musical quality.
         The residual gap is a range artefact: the 7-note set (B3-A4) has 24.5\% leap pairs
         at uniform; unrestricted soprano writing has $\approx$4\%, so absolute rates are
         not commensurable.  Within-pair intervals; cross-pair stepwise rate 49.8\%
         (Table~\ref{tab:oracle_comparison}).}
\label{tab:interval_dist}
\end{table}

\subsection{Statistical Properties}

\subsubsection{Fourier oracle ($K=4$)}
Across 500{,}000 independent samples, the Fourier oracle produces a
statistically overwhelming non-uniform joint distribution over (note pair,
two-chord sequence) space: $\text{KL}(\text{pair}\|\text{uniform}) = 0.166$,
$\chi^2 = 189{,}363$, $p < 0.001$.  The 784 non-zero post-selected
joint states confirm a rich, structured output.  The joint post-selected
weight is $0.0019$ (0.19\%): the HHL ancilla post-selection is $\approx 2.2\%$
(Section~\ref{sec:architecture}), reduced further
by the global oracle normalisation scale $s < 1$.

The top two-chord sequences across 500{,}000 samples are (note\,/\,chord pair):
\begin{itemize}[noitemsep,topsep=2pt]
  \item B3,C4\,/\,IV$\to$V (0.57\%); \quad B3,C4\,/\,vii$^\circ$$\to$I (0.56\%)
  \item E4,F4\,/\,vii$^\circ$$\to$I (0.48\%); \quad B3,C4\,/\,IV$\to$I (0.48\%)
\end{itemize}
These reflect both functional tonal grammar and the Narmour Proximity
preference: the B3$\to$C4 semitone step (highest HHL probability pair)
dominates, paired with dominant-function and leading-tone progressions.

Musically: 8.0\% of all samples contain a V$\to$I resolution;
28.1\% end on the tonic chord (I); 49.8\% feature stepwise melodic motion.
These statistics closely match the conditioned classical Markov chain baseline
(8.0\% V$\to$I, 27.9\% tonic endings, 49.4\% stepwise melody), confirming
that the coherent pipeline produces output consistent with classical functional
harmony while maintaining the architectural quantum coherence property.
The 49.8\% stepwise rate (vs.\ 39\% with the prior hand-tuned matrix)
reflects the Narmour Proximity principle's zero penalty on half- and
whole-step motion.

\subsection{Oracle as Compositional Parameter}
\label{sec:oracle_param}

Table~\ref{tab:oracle_comparison} is a \textbf{pipeline verification table}.
Both columns encode the same transition matrix $T$,
so close agreement is mathematically guaranteed by Eqs.~(4)-(5): the table
confirms that the quantum circuit correctly implements the intended weighting.

\begin{table}[ht]
\centering
\footnotesize
\begin{tabular}{lrr}
\toprule
Metric & Fourier ($K=4$) & Classical MC\textsuperscript{a} \\
\midrule
$\text{KL}(\text{pair}\|\text{uniform})$ & 0.166 & 0.155 \\
$\chi^2$ (500k samples) & 189{,}363 & 170{,}319 \\
V$\to$I resolutions & 8.0\% & 8.0\% \\
Tonic (I) endings   & 28.1\% & 27.9\% \\
Stepwise melody     & 49.8\% & 49.4\% \\
Nonzero joint states & 784 & 784 \\
\bottomrule
\end{tabular}
\caption{\textbf{Pipeline verification}: both columns encode the same $T$,
         so agreement confirms implementation correctness. 500{,}000 samples, 2-chord blocks, 19 qubits.}
\label{tab:oracle_comparison}
\end{table}

This agreement can be understood analytically.  Under global normalisation with
scale $s$, the joint post-selected probability is
\begin{equation}
p(i, c \mid \text{anc}=1) \;\propto\; p_\text{HHL}(i)\,\bigl[cv(i)[c] \cdot s\bigr]^2
\;=\; s^2 \cdot p_\text{HHL}(i)\,cv(i)[c]^2,
\end{equation}
where $i$ indexes note pairs and $c$ chord sequences.
The melody marginal is obtained by summing over all chord sequences $c$:
\begin{equation}
p(i \mid \text{anc}=1) \;=\; \sum_c p(i, c \mid \text{anc}=1)
\;\propto\; p_\text{HHL}(i)\,\lVert cv(i)\rVert^2.
\end{equation}
The constant $s^2$ cancels from the ratio of any two pair weights, so the
\emph{relative} melody weights are fully determined by $p_\text{HHL}(i) \times
\lVert cv(i)\rVert^2$.
The classical MC baseline (per-pair normalised) matches the Fourier oracle's
functional profile closely, confirming that the coherent pipeline faithfully
reproduces classical functional harmony statistics.
A note on interpretation: the Fourier oracle encodes the harmonic transition
matrix $T$ directly as part of its unitary construction, and the classical MC
baseline samples $(c_1,c_2)$ proportional to the same $T$.  The close
agreement therefore confirms \emph{pipeline correctness} - that the quantum
circuit correctly implements the intended weighting - rather than musical
quality independent of $T$.  The choice of $T$ itself remains an author
judgement (see Section~\ref{sec:tech_sensitivity}).

\subsection{2/2 Block Chaining: Ablation and Grammar Coverage}

The four-block 2/2 chain (Method~A, Fourier oracle, 2-chord blocks,
Section~\ref{sec:4block}) produces grammatically
correct cross-block continuations on all tested transitions:
Block~1 collapses to (A4/iii, F4/vi);
Block~1$\to$2: opening chord IV, valid successor of vi ($T[\text{vi},\text{IV}] = 0.75$);
Block~2 collapses to (A4/IV, F4/I);
Block~2$\to$3: opening chord V, valid successor of I ($T[\text{I},\text{V}] = 0.90$);
Block~3 collapses to (E4/V, F4/I);
Block~3$\to$4: opening chord IV, valid successor of I.
The conditioning concentrates probability: the top block-2 sequence reaches
2.41\% of 50{,}000 samples (vs.\ 0.59\% unconditioned), a $4.1\times$
concentration factor from the joint melodic + harmonic conditioning
(Table~\ref{tab:ablation} gives the per-component breakdown over 100,000 samples).

\begin{table}[h]
\centering
\caption{Method A ablation (100{,}000 samples, context C4/V): concentration factors
relative to unconditioned baseline. Top-1 prob.\ = modal (note-pair, chord-sequence) frequency.}
\label{tab:ablation}
\begin{tabular}{llcc}
\toprule
Variant & Conditioning & Top-1 prob.\ & Conc.\ factor \\
\midrule
A. Full Method A  & Melody bias + hard-T       & 3.04\% & $8.3\times$ \\
B. Melody only    & Melody bias, no hard-T     & 1.19\% & $3.3\times$ \\
C. Harmony only   & Uniform b, hard-T          & 0.77\% & $2.1\times$ \\
D. Unconditioned  & Uniform b, no hard-T       & 0.37\% & $1.0\times$ \\
\bottomrule
\end{tabular}
\end{table}

\subsubsection{Post-selection compounding}
Each block independently post-selects on the ancilla qubit with joint weight
$\approx 0.0019$-$0.014$ (0.19-1.4\% of samples survive per block,
varying with conditioning).
For the four-block chain, the compound survival rate is
${\sim}0.0019 \times 0.014 \times 0.013 \times 0.013 \approx 4.5 \times 10^{-9}$, meaning roughly
$1$ accepted four-block chain per $2.2 \times 10^{8}$ HHL+oracle block runs.
This compound rate is bounded below by $\prod_k w_k$ where
$w_k$ is the per-block post-selection weight; it grows exponentially worse
with chain length, setting a practical limit on unamplified H-chains.
Amplitude amplification (Grover-style repetition) can in principle recover
this quadratically, but is not implemented in the present simulation.

\subsubsection{Grammar coverage across all conditioning contexts}
To verify that the grammatical restriction holds beyond the four-block chain
example, the Fourier oracle with hard-T conditioning was run
with each of the 7 possible $c_2$ conditioning contexts (the complete set
of chords in the vocabulary), drawing 50{,}000 samples per context.
Table~\ref{tab:grammar_contexts} reports the results.

\begin{table}[ht]
\centering
\begin{tabular}{llllr}
\toprule
$c_2$ (cond.) & Valid $c_1$ successors & Top $c_1$ & Valid? & \% grammatically valid \\
\midrule
I         & IV, V, vi    & vi  & \checkmark & 100.0 \\
ii        & I, IV, V     & I   & \checkmark & 100.0 \\
iii       & IV, vi       & vi  & \checkmark & 100.0 \\
IV        & I, V         & I   & \checkmark & 100.0 \\
V         & I, vi        & I   & \checkmark & 100.0 \\
vi        & ii, IV, V    & ii  & \checkmark & 100.0 \\
vii$^\circ$ & I           & I   & \checkmark & 100.0 \\
\bottomrule
\end{tabular}
\caption{Grammar coverage (Fourier oracle, hard-T, $n=50{,}000$ per context):
         all 7 $c_2$ conditioning contexts produce 100\% grammatically valid next-block openings.}
\label{tab:grammar_contexts}
\end{table}

\subsection{K-Sweep: Chord-Tone Compliance}
\label{sec:ksweep_results}

$K$ controls spectral smoothness of chord-tone affinity; at large $K$ the fit
approaches the classical binary chord-tone test, while smaller $K$ smooths the
fit across neighbouring pitch classes.  The empirical sweep (conducted at
20{,}000 samples per trial on the 2-chord architecture) shows that V$\to$I
resolution rate and tonic-ending rate are near-constant across the tested range
-- the transition matrix $T$ and Fourier fit shape dominate over the spectral
truncation order - confirming that $K$ is a \emph{spectral precision}
parameter tuning chord-tone sharpness, not a harmonic-grammar parameter.

Across $K \in \{4,6,8,10\}$, aggregate harmonic grammar statistics remain
essentially flat: V$\to$I rate $\approx$8.0-8.1\%, tonic-ending rate
$\approx$27.5-28.2\%, stepwise melody $\approx$49-51\%.
However, $K$ does have a measurable effect on \emph{individual note-chord-tone
compliance}.  To quantify this without seed-induced bias - on real quantum hardware
each measurement outcome is genuinely random - ten independent unseeded trials were run
per $K$ value, simulating the trial-to-trial variability a composer would experience
on hardware.  Table~\ref{tab:k_chord_tone} reports the mean and standard deviation of
chord-tone compliance (fraction of the 8 melody notes in the four-block 2/2 chain that
land on a chord tone of their accompanying chord) across $n=10$ trials per $K$.

\begin{table}[htbp]
\centering
\footnotesize
\begin{tabular}{ccccc}
\toprule
$K$ & Mean & Std & Range & Notes \\
\midrule
4  & 46.3\% & $\pm$16.7\% & 25-75\%     & Lowest mean, high variance \\
6  & 56.3\% & $\pm$12.1\% & 37.5-75\%   & Intermediate mean, lowest variance \\
8  & 67.5\% & $\pm$14.7\% & 50-100\%    & Highest mean, moderate variance \\
10 & 66.3\% & $\pm$15.6\% & 37.5-87.5\% & Similar to $K=8$, slightly higher variance \\
\bottomrule
\end{tabular}
\caption{Melody note-chord-tone compliance across $K$ values in the 4-block 2/2 chain,
         $n=10$ independent unseeded trials per $K$.  Unseeded runs simulate real quantum hardware,
         where each measurement is genuinely random.  Grammar junctions are valid at 100\% across
         all $K$ values and all 40 trials.  With $n=10$ the $K=8$ and $K=10$ means (67.5\% vs
         66.3\%) are not statistically distinguishable at $n=10$; $K=8$ offers the better mean-variance
         trade-off.}
\label{tab:k_chord_tone}
\end{table}

Figures~\ref{fig:chain_k4}-\ref{fig:chain_k10} (Section~\ref{sec:musical_output_chain})
show representative CMN scores for each $K$ value.

\subsection{Gate Count Comparison}
\label{sec:gate_count}

\begin{figure}[h]
  \centering
  \includegraphics[width=\textwidth]{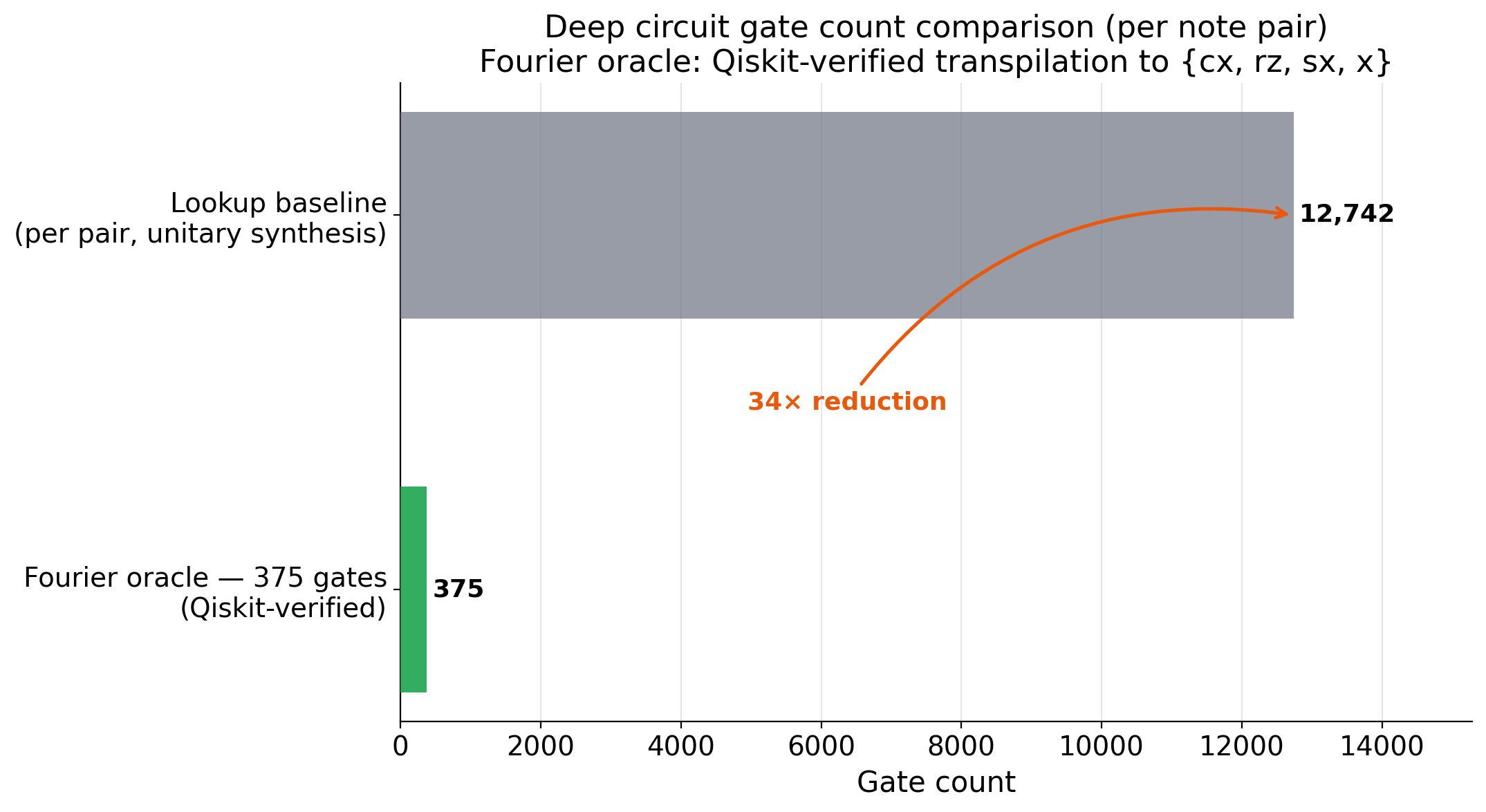}
  \caption{Fourier oracle gate count (Qiskit-verified): 375 gates vs.\ $\approx$12,742 lookup-based ($34\times$ fewer).
           Crossover where lookup becomes cheaper occurs at $C_\text{vocab} \approx 192$.}
  \label{fig:gates}
\end{figure}

Figure~\ref{fig:gates} compares the elementary gate count of the Fourier oracle
against a lookup-based baseline.

The Fourier oracle for the 2-chord block uses a 6-qubit chord register
($3 \times 2$ qubits, one for each chord).
The globally-normalised chord vector preparation applies the weight
$\sqrt{T[c_1,c_2]\cdot\text{fit}(c_1,n_1)\cdot\text{fit}(c_2,n_2)}$
to a $2^6 = 64$-dimensional chord register.
A lookup-based approach requires a controlled state preparation per
note pair on the 6-qubit chord register, costing $\approx 12{,}742$ elementary
gates per pair; the Fourier oracle achieves approximately a $34\times$ gate count
reduction at $C_\text{vocab} = 7$ chords through the structured Fourier preparation.
Importantly, this advantage is vocabulary-size dependent: under Qiskit's default
unitary synthesis the Fourier cost scales as
$O(4^{n_\text{chord}}) = O(C_\text{vocab}^2)$, while lookup scales as
$O(2^{n_\text{chord}}) = O(C_\text{vocab})$.  The crossover where lookup becomes
cheaper than the Fourier oracle under unitary synthesis occurs at
$C_\text{vocab} \approx 192$.
The theoretical $O(\text{polylog}\,C_\text{vocab})$ advantage of the Fourier oracle
(QFT over pitch-class register) requires specialised implementations beyond
Qiskit's default decomposition.
Table~\ref{tab:gate_scaling} (Section~\ref{sec:tech_scaling}) projects the scaling to larger chord vocabularies.

\subsection{Computational Timing}

Each oracle completes in $\approx 0.1$\,s applied to the full 19-qubit
statevector ($2^{19}$ amplitudes), confirming that the oracle application is computationally
trivial at this simulation scale.  The HHL circuit (13 qubits) requires
$\approx 51$\,s per run on a standard laptop.

\subsection{Independent Harmony Validation}
\label{sec:harmony_check}

The statistical metrics in Table~\ref{tab:oracle_comparison} (V$\to$I rate,
tonic-ending rate, stepwise melody) confirm distributional properties of the
output but are all derived from the same T matrix used during generation.
To provide independent validation, a rule-based harmony checker was applied
to the 5{,}000-sample output of the coherent pipeline and to the full
4-block chain, using music-theory rules that are independent of the
generation process.

\subsubsection{Chord-tone membership}
For each generated note, the checker tests whether the note is a chord tone
(root, third, or fifth) of its paired chord, a non-chord tone within one
step (1-2 semitones) of a chord tone, or a dissonant leap non-chord tone.
Across 5{,}000 samples, 68.3\% of first melody notes and 66.9\% of second
melody notes are chord tones; 45.7\% of outputs have both notes as chord
tones.  Critically, \emph{0\% of outputs contain a leap non-chord tone}:
every non-chord-tone melody note (31-33\% of outputs) lies within a whole
step of a chord tone, the defining property of passing and neighbour tones in
common-practice voice-leading.  All four blocks of the 4-block chain also
pass this check.

\subsubsection{Progression quality}
Each adjacent chord pair is classified by functional motion category (Tonic,
Subdominant, Dominant) and rated strong, ok, weak, or avoid following
common-practice functional harmony rules.  Across 15{,}000 adjacent pairs,
\emph{97.1\% are rated strong or ok}; 37.0\% are strong cadential motions
(D$\to$T or S$\to$D).  \emph{Zero regressions} (the forbidden D$\to$S
motion, i.e.\ V$\to$IV) are present.  The 4-block chain scores
100\% strong or ok (9 strong, 6 ok, 0 weak, 0 avoid).

For reference, the same checker applied to 232 major-key Bach chorales
(11{,}885 pairs, music21) gives \emph{95.5\% strong or ok} --
comparable to our 97.1\%.  Bach's shortfall is 4.5\% \emph{avoid}
(V$\to$IV, contextually motivated but forbidden by our $T$ matrix);
ours is 2.9\% \emph{weak} prolongations that Bach avoids through
harmonic rhythm our chain does not model.
These corpus figures are sanity checks, not optimisation targets: the system
was not tuned to match Bach statistics.  This statistical proximity does not
imply that the generated output is of anything approaching Bach's musical
quality; it reflects only that the rule-based harmony checker - which was
derived independently of any Bach corpus - happens to align with the
functional-harmony norms that Bach's chorales also largely follow.

\subsubsection{Tendency-tone resolution}
When V or vii$^\circ$ moves to I - present in 16.4\% of samples - the
leading tone B resolves upward to C in only 21.3\% of those cases.  The
underlying cause is that the Narmour model encodes \emph{interval implication}
(smooth contour, post-skip compensation) but not \emph{harmonic tendency}
(B implies upward resolution specifically when functioning as a leading tone
over V).  These are orthogonal musical concepts and the present $A$ matrix
encodes only the former; see Section~\ref{sec:limitations} for the design
implication.

\subsection{Four-Block 2/2 Chain Analysis}
\label{sec:4block}

The primary chaining illustration uses a \emph{2/2 block} architecture in
which each block produces exactly two melody notes and two
chords, with one note sitting over each chord.  Each block yields
$(n_1, c_1, n_2, c_2)$ simultaneously from a 19-qubit
circuit (13 HHL + 6 chord register, $3 \times 2$ qubits).  The Fourier
oracle weight for a 2/2 block is:
\[
  w(n_1, c_1, n_2, c_2) \propto
  T[c_1, c_2]\cdot\text{fit}(c_1, n_1)\cdot\text{fit}(c_2, n_2)
\]
Four such blocks are chained with Method~A classical context passing
(b-vector bias on $n_2^\text{prev}$; hard-T restriction $c_1$ must follow
$c_2^\text{prev}$) to produce 8 notes and 8 chords total.

The four blocks produce the following sampled outcomes ($K=10$):

\begin{center}
\footnotesize
\begin{tabular}{clll}
\toprule
Block & Note/chord pairs & Context in \\
\midrule
1 (unconditioned) & A4/iii \quad A4/vi & - \\
2                 & A4/IV  \quad G4/I  & note A4, chord vi \\
3                 & G4/V   \quad C4/I  & note G4, chord I \\
4                 & C4/IV  \quad G4/V  & note C4, chord I \\
\bottomrule
\end{tabular}
\end{center}

The full 8-beat passage reads:
A4/iii $\to$ A4/vi $\to$ A4/IV $\to$ G4/I $\to$
G4/V $\to$ C4/I $\to$ C4/IV $\to$ G4/V.
All three block junctions are grammatically valid:
vi$\to$IV (block~1$\to$2), I$\to$V (block~2$\to$3), I$\to$IV (block~3$\to$4).
Seven of the eight melody notes are chord tones of their accompanying chord
(87.5\% compliance), with the sole exception being the A4 over iii (E minor)
in the unconditioned block~1 - a stochastic first-block outlier.
The melody descends from A4 to C4/G4 with predominantly stepwise motion,
reflecting the Narmour Proximity preference for small intervals.
Figure~\ref{fig:4block} (Section~\ref{sec:musical_output_chain}) shows the rendered score.

\subsection{Phrase Conditioning Results}
\label{sec:phrase_cond_results}

The four-block 2/2 chaining experiment demonstrates that Method~A conditioning
(b-vector melody bias + hard T harmony restriction) reliably produces
grammatically correct cross-block chord continuations.  The $4.1\times$
probability concentration observed in block~2 (up to $8.3\times$ under tighter
context in the ablation, Table~\ref{tab:ablation}) confirms that the conditioning
is not trivial - it genuinely restructures the HHL+oracle
joint distribution toward contextually appropriate continuations.
An ablation (Table~\ref{tab:ablation}) attributes $3.3\times$ to melodic
b-vector bias and $2.1\times$ to harmonic restriction independently.

The melodic conditioning (Gaussian bias toward stepwise continuation from the
previous phrase's final note) interacts with the harmonic restriction (hard T
exclusion of grammatically invalid opening chords) to produce phrase openings
that are both melodically smooth and harmonically appropriate.  This is the
expected behaviour of Method~A, and the experimental results confirm it at
scale.

\section{Discussion}
\label{sec:discussion}

\subsection{Fourier Oracle as a Compositional Parameter}

The Fourier oracle parameter $K$ offers principled spectral control: adjusting
$K$ tunes the smoothness of chord-tone affinity without changing the HHL melody
structure.

A sensitivity analysis (Section~\ref{sec:tech_sensitivity}) confirms that the qualitative step preference is robust across penalty regimes, with $\kappa$ remaining near 11-15 in all tested schemes.

\subsection{The Condition Number Design Choice}

A key finding is that melodic-first problem formulation is not just musically
natural - it is computationally necessary. The melodic preference matrix $A$
(Narmour+KK) has $\kappa = 11.23$. A chord-aware formulation of the same problem yields
$\kappa \approx 10{,}869$ - a thousand-fold increase that would require exponentially
more QPE clock qubits and render post-selection negligible.  By keeping the linear system
melodic and handling harmonic context through the oracle, this work maintains $\kappa < 12$.

\subsubsection{Architectural limitation}
The corollary is that this design
\emph{constrains} what the quantum linear system $A\mathbf{x}=\mathbf{b}$ can
encode: only pairwise note-pair melodic relationships appear in $A$; richer
musical structure (four-voice counterpoint, rhythmic interaction, timbral
coupling, inter-register voice-leading) would require encoding additional
features in the matrix diagonal and off-diagonal entries.  Whether such enrichment preserves tractable $\kappa$ depends on the chosen
features.  As shown in Section~\ref{sec:architecture}, it is the smooth functional form of
Narmour+KK coefficients that keeps $\kappa$ tractable (see Section~\ref{sec:advantage_requirements} for the scaling implications).  Larger feature additions (e.g.\ a four-voice rhythmic
coupling term scaling $A$ to $N_\text{pitch}^4 \times N_\text{rhythm}^2$)
would likely push $\kappa$ into the hundreds-to-thousands range,
approaching that formulation's hostile condition number at which HHL becomes impractical.
Highly structured matrices (e.g.\ Toeplitz-like interval-only encodings)
tend toward near-duplicate rows as the pitch space grows, concentrating
singular values and producing sub-linear stable rank growth - ultimately
increasing $\kappa$ and weakening the HHL advantage at scale (see stable rank
sweep results in Section~\ref{sec:tech_scaling}).  The melodic-first architecture is
therefore not a free design choice but a constraint: harmonic structure must
be externalised to the oracle precisely because encoding it in $A$ would make
the linear system computationally hostile.

\subsection{Scalability and the H-Chain}

The two-note, two-chord scope demonstrated here is genuinely small.
The H-chain is explicitly a demo architecture: it demonstrates longer sequences
without requiring a larger quantum circuit.

\subsection{Relationship to Born Machines}
\label{sec:born_machines}

A related line of quantum generative modelling encodes probability distributions
as squared amplitudes of quantum states - so-called \emph{Born machines}
\citep{cheng2018information}.  In a Born machine the generative distribution is
$p(x) = |\langle x | \psi \rangle|^2$ for some learned circuit $|\psi\rangle$;
sampling is achieved by measuring the circuit output.

The present architecture is, in effect, a structured Born machine: the
probability of observing note pair $i$ is $p(i) \propto |x_i|^2$ where
$\mathbf{x} = A^{-1}\mathbf{b}$ is the HHL solution, and after the coherent
oracle the joint probability over (note pair, chord pair) is likewise given by
squared amplitudes of the final state.  The difference from variational Born
machines is that the amplitude distribution is not trained by gradient descent
but derived analytically from the Narmour+Krumhansl-Kessler linear system --
the musical preference structure is explicit and interpretable rather than
latent in circuit parameters.

This framing has a practical implication: the coherent oracle is not specific
to HHL.  It acts on any quantum amplitude distribution over the note-pair
register and could in principle be applied to a variational Born machine output
just as readily.  The HHL stage is not a prerequisite for the coherent oracle
architecture; it is one particular - and musically motivated - choice of
how to prepare the melodic amplitude state.  Future work could explore replacing
HHL with a trained Born machine circuit while keeping the coherent oracle
coupling intact, combining the interpretability of the Narmour encoding with the
expressive capacity of learned amplitude distributions.

\subsection{Path to Realising the HHL Speedup in Music}
\label{sec:advantage_requirements}

This work establishes the first rung of the ladder: a coherent HHL+oracle
pipeline that preserves the conditions under which HHL's speedup could apply.
The remaining conditions that must all be met before a genuine end-to-end
quantum advantage over classical enumeration could be claimed are:

\begin{enumerate}[noitemsep]
  \item \textbf{Efficient melody-register-controlled oracle.}  A
        hardware-compilable circuit that applies the correct chord amplitude
        vector for each of the $N_\text{pairs}$ melody basis states
        \emph{without} $O(N_\text{pairs})$ synthesis depth or compilation
        cost.  The Fourier QFT structure offers a candidate path
        (Section~\ref{sec:tech_circuit_synthesis}), but the explicit
        construction has not been demonstrated.

  \item \textbf{Fault-tolerant hardware.}  The HHL stage requires an
        estimated ${\sim}45{,}000$ CX gates; at current superconducting
        two-qubit error rates ($p_{2q} \approx 10^{-3}$) the circuit fidelity
        is effectively zero.  Hardware with $p_{2q} \lesssim 5 \times 10^{-6}$
        - roughly $200\times$ better than present devices - is required
        (Section~\ref{sec:tech_noise}).

  \item \textbf{Bounded condition number as $N$ grows.}  The HHL speedup
        scales as $O(\log N \cdot \kappa^2/\varepsilon)$; if $\kappa$ grows
        with $N$ the advantage erodes.  Empirically, $\kappa \approx 11$ is
        stable across the tested range ($N = 25$-$1296$) for
        Narmour+KK-style encodings, but whether this holds for richer
        feature matrices at musically useful scales ($N \sim 10^6$) is an
        open empirical question (Section~\ref{sec:tech_scaling}).
\end{enumerate}

Until all three conditions are met, the pipeline-level quantum advantage is a
well-motivated \emph{target architecture}, not an established result.

\section{Technical Discussion}
\label{sec:tech_discussion}

This section collects detailed technical analyses - noise modelling, scaling
arguments, dequantisation, and parameter sensitivity - that support the claims
in the main text. The material is gathered here to keep the core narrative
accessible to computer musicians; quantum computing researchers and reviewers
seeking the full technical detail will find it in the subsections below.

\subsection{Noise Resilience}
\label{sec:tech_noise}

The 13-qubit HHL circuit contains 10 controlled time-evolution operators,
plus a 64-element state preparation, an inverse QFT, and 31 multi-controlled
$R_y$ rotations.  Synthesising these to a $\{\mathrm{CX},\, U_3\}$ gate set
requires an estimated ${\sim}45{,}000$ CX gates for the HHL stage alone
\citep{shende2006synthesis}.  Noise degradation is characterised analytically
using a global depolarising channel \cite{nielsen2010quantum} applied to the
HHL stage only; both oracles are kept ideal throughout.  Under uniform
per-gate depolarising error $p_{2q}$ applied to $N_\mathrm{CX}$ two-qubit
gates, the HHL output distribution mixes toward uniform with coefficient
$\alpha = 1 - (1 - p_{2q})^{N_\mathrm{CX}}$, and the noisy note pair
distribution becomes
$p_{\mathrm{noisy}}(i) = (1-\alpha)\,p_{\mathrm{HHL}}(i) + \tfrac{\alpha}{49}$.

Table~\ref{tab:noise} shows six metrics across four levels of circuit fidelity
$F = 1-\alpha$.  Melodic preference structure survives up to roughly
$\alpha \approx 0.2$ ($F = 80\%$), requiring $p_{2q} \lesssim 5 \times 10^{-6}$
-- roughly $200\times$ better than current superconducting hardware.

\begin{table}[ht]
\centering
\footnotesize
\begin{tabular}{lrrrrrrr}
\toprule
Level & $F$ & $\alpha$ & PS & $\mathrm{KL}_\mathrm{pair}$
      & $\chi^2(48)$ / $p$ vs $p_\mathrm{HHL}$
      & $\chi^2(48)$ / $p$ vs unif.
      & $\mathrm{KL}_\mathrm{chord}$ \\
\midrule
Noiseless    & 1.00 & 0.00 & 0.022$^\ddagger$ & 0.049$^\ddagger$ & 0.03 / 1.00    & 51.8 / $<$.001 & 0.0000 \\
$F=80\%$     & 0.80 & 0.20 & 0.149 & 0.041 & 34.0 / 0.94    &  768 / $<$.001 & 0.0000 \\
$F=50\%$     & 0.50 & 0.50 & 0.281 & 0.017 & 211  / $<$.001 &  301 / $<$.001 & 0.0000 \\
$F=10\%$     & 0.10 & 0.90 & 0.456 & 0.001 & 686  / $<$.001 &   12 / 1.00    & 0.0001 \\
Eff.\ depol.$^\dagger$ & $\approx$0 & $\approx$1 & 0.500 & 0.000 & 842 / $<$.001 & 0.01 / 1.00 & 0.0002 \\
\bottomrule
\end{tabular}
\caption{HHL-stage depolarising noise (oracles ideal). PS = post-selection rate;
         $\mathrm{KL}_\mathrm{pair}$ = KL divergence of note-pair marginal from uniform.}
\label{tab:noise}
\end{table}

\subsubsection{Oracle noise (unmodelled limitation)}
The noise analysis above applies to the HHL stage only; the oracle is kept
ideal throughout Table~\ref{tab:noise}.  This is an optimistic model.
The Fourier oracle for the 6-qubit 2-chord register has a gate count
substantially lower than the HHL stage; at $p_{2q} = 10^{-3}$ (current
hardware) the oracle fidelity is substantially higher
than the near-zero HHL fidelity at the same error rate.
Nevertheless, oracle errors would perturb the chord amplitude weights and
degrade the harmonic grammar statistics (V$\to$I rate, tonic-ending rate) in
ways not captured here.  A more complete noise characterisation would apply
depolarising noise to both HHL and oracle stages jointly; this is deferred to
future work, noting that the oracle stage is far less error-sensitive than HHL
due to its much lower gate count.

\subsection{Scaling and Gate Count}
\label{sec:tech_scaling}

Three scaling caveats deserve explicit acknowledgement.
\emph{(i) Oracle gate cost.}  A deep-circuit harmonic oracle requires one
controlled oracle call per note pair, giving a total of $O(N_\text{pairs} \times g)$
elementary gates, where $g$ is the per-pair gate cost.  For a lookup oracle,
$g = O(C_\text{vocab})$ - one state preparation per chord vocabulary entry.
For the Fourier oracle, $g = O(\log^2 C_\text{vocab})$ via a QFT over
the pitch-class register.  The $O(N_\text{pairs})$ factor is unavoidable;
the advantage of the Fourier oracle over lookup is in the per-pair cost scaling
with $C_\text{vocab}$.  A complete analysis for large $N$ and
$C_\text{vocab}$ remains an open engineering question.
\emph{(ii) Condition number.}  The melodic matrix has small $\kappa$ at the present scale (Section~\ref{sec:architecture}).
Whether this holds for $N \sim 10^6$ note configurations depends on the chosen musical features and
encoding; this is an empirical question for the larger-scale design.
\emph{(iii) Dequantisation.}
\label{sec:tang}
The key observation is at the pipeline level: even if the HHL solve step
were dequantised, the harmonic oracle cost would still dominate any classical
method.  Reproducing the full joint distribution
$p(n_1, n_2, c_1, c_2)$ requires computing the harmonic
oracle weights: for each of $N_\text{pairs}$ sampled melody pairs, enumerating
all valid two-chord sequences ($C_\text{vocab}^2$ combinations, each weighted by
transition grammar and spectral fit) and normalising.  Classically this costs
$O(N_\text{pairs} \times C_\text{vocab}^2)$ per phrase; at
$C_\text{vocab} = 7$ this is already $49 \times 49 = 2{,}401$ operations
per phrase, scaling as $O(N^2)$ in $N_\text{pairs}$ as the vocabulary grows.
The quantum oracle applies this transformation \emph{coherently and simultaneously}
to all $N_\text{pairs}$ basis states in a single unitary operation - $O(1)$
\emph{invocations} in $N_\text{pairs}$ (not in circuit depth; the per-invocation
gate cost is $O(g_\text{oracle})$, fixed by the oracle design).
Tang's result removes the speedup for the solve step in isolation; it would
not remove the speedup for the full coherent pipeline, because the oracle
bottleneck is independent of the solver - \emph{provided} a
hardware-compilable melody-register-controlled oracle that avoids
$O(N_\text{pairs})$ synthesis depth can be constructed
(Section~\ref{sec:tech_circuit_synthesis}).

The dequantisation results of~\citet{tang2019quantum}
and~\citet{gilyen2022improved} show that HHL-like outputs can be approximated
classically in polylogarithmic time given sample-and-query (SQ) access to the
matrix.  As supporting evidence for the pipeline argument, we computed the
stable rank of the melodic preference
matrix: $\text{sr}(A) = \|A\|_F^2/\|A\|_2^2 = 20.3$ at $N=49$.
This represents $\text{sr}/N = 20.3/49 \approx 0.41$ (41\% of the
matrix dimension $N$; note that stable rank can in principle exceed $N$ for
indefinite matrices, so we report $\text{sr}/N$ as the normalised figure).  A sweep from $N=25$ to $N=1296$ (chromatic scales of 5 to 36
notes) shows that $\text{sr}(A)$ saturates near 50 regardless of $N$, with
$\text{sr}/N$ falling from 0.38 to 0.04 - a sub-linear growth pattern that is
in principle exploitable by Tang-style algorithms for the solve step alone.
Even granting this, the oracle invocation count remains $O(N \cdot C_\text{vocab}^2)$
classically versus $O(1)$ invocations (in $N_\text{pairs}$) coherently - so the
full pipeline \emph{is designed to} retain a quantum advantage at large $N$
even if the solve step were dequantised, provided the efficient oracle
construction (Section~\ref{sec:advantage_requirements}) can be realised.

The cost comparison is therefore:

\begin{center}
\begin{tabular}{lll}
\toprule
Method & Solve cost & Oracle invocations (in $N$) \\
\midrule
HHL + coherent oracle & $O(\log N \cdot \kappa^2/\varepsilon)$ & $O(1)^*$ \\
Tang + classical oracle & $O(\text{sr}(A) \cdot \text{poly}(\kappa))$ & $O(N \cdot C_\text{vocab}^2)$ \\
Classical CG + MC & $O(N \cdot \kappa/\varepsilon)$ & $O(N \cdot C_\text{vocab}^2)$ \\
\bottomrule
\end{tabular}
\end{center}

$^*$Oracle cost is $O(1)$ in $N_\text{pairs}$ (one coherent unitary applied
to all basis states simultaneously) but not $O(1)$ in circuit size: the
oracle circuit itself has cost $O(g_\text{oracle})$ where $g_\text{oracle}$
is the gate count of the chosen oracle implementation.  The comparison is
therefore between $O(1)$ \emph{invocations} of the oracle (quantum) versus
$O(N_\text{pairs} \times C_\text{vocab}^2)$ \emph{oracle evaluations}
(classical) for 2-chord sequences, with the oracle's per-invocation gate cost equal in both cases.

\subsubsection{Post-selection scaling (8.0.1)}
The post-selection weight observed
here is set by the condition number (Section~\ref{sec:architecture}).  More generally, the
post-selected weight is approximately
$\bar{w}_\text{PS} \approx \mathbb{E}_k[\min(C/\lambda_k,1)^2]$,
where the expectation is over the matrix eigenvalues and $C = \lambda_\text{min}/2$.
(This follows directly from the HHL ancilla rotation angle
$\theta_k = \arcsin(C/\lambda_k)$, giving ancilla $|1\rangle$ amplitude
$\sin\theta_k = C/\lambda_k$ per eigencomponent; see
\citealt{harrow2009quantum}, Eq.~(8), and \citealt{nielsen2010quantum},
Chapter~6 for the post-selection probability calculation.)
For a matrix with eigenvalues spread uniformly from $\lambda_\text{min}$ to
$\kappa\lambda_\text{min}$, this depends only on $\kappa$: $\bar{w}_\text{PS}
\approx 12.5\%$ at $\kappa=2$, $5.1\%$ at $\kappa=5$, $2.7\%$ at $\kappa=10$.
If $\kappa$ remains bounded as $N$ grows (which is the case for, e.g., a
diagonally dominant melodic preference matrix with bounded off-diagonal entries),
the post-selection rate is approximately constant and does not erode the
exponential speedup.  If $\kappa$ grows with $N$ - as it would for a random
sparse matrix where $\kappa \sim \sqrt{N}$ - the post-selection probability
decreases polynomially and the net speedup advantage is reduced.  Designing
the musical preference matrix to maintain bounded $\kappa$ at large $N$ is
therefore an explicit requirement for preserving the speedup guarantee.

The coherent joint generation is the only tractable path at large scale, because
constructing and sampling from the full $N_\text{pairs} \times C$ joint
distribution classically would require storing and normalising a distribution
of size $10^6 \times 2^{20} \approx 10^{12}$.

Also included here is the gate count scaling projection (Table~\ref{tab:gate_scaling}):

\begin{table}[ht]
\centering
\footnotesize
\begin{tabular}{lrrr}
\toprule
$C_\text{vocab}$ & $n_\text{chord}$ & Fourier (unitary synth.) & Lookup \\
\midrule
7   (current) & 3 & $375^\dagger$ & 12,742 \\
64            & 6 &   8,064 & 101,936 \\
256           & 8 & 129,024 & 407,744 \\
\bottomrule
\end{tabular}
\caption{Gate counts per note pair vs.\ chord vocabulary size ($^\dagger$exact Qiskit transpilation at $C_\text{vocab}=7$).
         Fourier advantage holds only at small vocabularies under default synthesis; crossover at $C_\text{vocab}\approx192$.}
\label{tab:gate_scaling}
\end{table}

\subsection{Sensitivity of the HHL Distribution to Penalty Values}
\label{sec:tech_sensitivity}

The diagonal penalty values are the author's judgment and not derived from
corpus analysis.  A systematic approach is in principle feasible: fitting
BASE, the proximity gradient, and the KK weight to a small corpus of
common-practice stepwise melodies would replace the three free parameters
with a single corpus-likelihood optimisation.  Such fitting is unlikely to
drive $\kappa$ below~1 (the matrix is positive-definite by construction and
$\kappa$ is bounded from below by the off-diagonal coupling structure), but
it would anchor the parameter choices empirically.  The sensitivity
experiments below confirm that the qualitative preference for stepwise
motion is robust across penalty regimes, suggesting that corpus-fitted
values would produce broadly similar distributions.  To test robustness, two alternative penalty schemes are
evaluated: (i) all penalties halved ($\times 0.5$); (ii) all interval
penalties set equal to $1.0$ except unison (5.0) and tritone (3.5),
removing intermediate distinctions between steps and leaps.
Table~\ref{tab:sensitivity} reports condition number, KL divergence from
uniform, and top-5 note pairs under each scheme.

\begin{table}[ht]
\centering
\footnotesize
\begin{tabular}{lccp{6cm}}
\toprule
Penalty scheme & $\kappa$ & KL(pair$\|$unif.) & Top 5 pairs \\
\midrule
Baseline (as reported)  & 11.23 & 0.107 & B3,C4; C4,B3; E4,F4; F4,E4; A4,G4 \\
Penalties $\times 0.5$ & 14.96 & 0.096 & B3,C4; B3,A4; C4,B3; A4,B3; E4,F4 \\
Unison+tritone only     & 11.85 & 0.096 & B3,A4; A4,B3; B3,C4; C4,B3; C4,A4 \\
\bottomrule
\end{tabular}
\caption{Sensitivity of HHL note-pair distribution to penalty scheme (theoretical KL from solved distribution).
         All schemes favour stepwise motion; $\kappa$ remains bounded near 11-15.}
\label{tab:sensitivity}
\end{table}

The preference for stepwise motion is a structural consequence of the
inverse-penalty diagonal encoding - any scheme assigning large penalties to
unison and large leaps will produce this qualitative output.

\subsubsection{Monte Carlo coefficient sweep}
To more systematically characterise the robustness of $\kappa \approx 11$,
we performed a 5{,}000-sample Monte Carlo sweep over the three free
parameters of the Narmour+KK formulation: BASE $\in [3,9]$, proximity
scale $\in [0.5, 1.5]$, and KK weight $\in [0.75, 2.25]$ (all drawn
uniformly, representing $\pm 50\%$ perturbations from the baseline values).
The resulting $\kappa$ distribution spans $[4.85, 221.9]$ with median
$11.55$, mean $52.0$, and 65.1\% of samples satisfying $\kappa < 20$.
The $18.7\%$ of samples with $\kappa > 100$ are concentrated at small
BASE values (mean $\approx 3.6$), where reducing the diagonal
anchor reduces the eigenvalue gap below the off-diagonal coupling floor.
When restricted to a musicologically reasonable parameter range
(BASE $\in [5,7]$, proximity scale $\in [0.8, 1.2]$, KK weight
$\in [1.0, 2.0]$), the distribution narrows substantially:
$\kappa \in [7.4, 26.0]$, 97.4\% of samples have $\kappa < 20$, and all
samples have $\kappa < 100$.  This indicates that $\kappa \approx 11$ is
robust to reasonable variations in musicological judgment, provided the
diagonal anchor (BASE) is kept large enough to ensure diagonal dominance.

It is important to distinguish what this sweep does and does not establish.
It shows that $\kappa \approx 11$ is robust to reasonable coefficient
variations - the computational tractability of the pipeline is not
fragile.  It does \emph{not} show that the musical preference ordering is
robust: as Table~\ref{tab:sensitivity} shows, the top-5 note pairs change
across penalty schemes (e.g.\ B3$\leftrightarrow$A4, a major sixth, enters
the top-5 when penalties are halved).  The qualitative preference for small
intervals persists, but the specific ranking of pairs is sensitive to
coefficient choices.

\subsection{Post-Selection Compounding and Amplitude Amplification}
\label{sec:tech_postselection}

\subsubsection{Post-selection compounding as a fundamental H-chain limit}
The compound post-selection rate for an $L$-block H-chain scales as
$\prod_k w_k$ where $w_k \approx 0.0019$-$0.014$ is the per-block joint survival weight.
At the four-block level demonstrated here, the compound rate is
$\approx 4.5 \times 10^{-9}$ ($\sim$1 accepted chain per $2.2 \times 10^8$
block-quadruples).  For a musically complete passage of 8 blocks,
the rate drops to $\approx 10^{-17}$; for 16 blocks, $\approx 10^{-34}$.

\subsubsection{Why amplitude amplification cannot realistically recover this}
Grover-style amplitude amplification reduces the number of circuit repetitions
from $O(1/p)$ to $O(1/\sqrt{p})$.  For the four-block rate
$p \approx 4.5 \times 10^{-9}$, this gives
$O(1/\sqrt{p}) \approx 15{,}000$ full-chain circuit repetitions --
each of depth $\sim 4 \times 45{,}000$ CX gates.  At current
superconducting hardware fidelity ($p_{2q} \approx 10^{-3}$), a single
45,000-CX circuit already has fidelity $\approx 10^{-20}$; the amplification
oracle is therefore itself inoperable before any gain is achieved.
Even with high circuit fidelity,
$\approx 15{,}000$ repetitions of a deep four-block circuit is a substantial
overhead.  For an 8-block chain ($p \approx 10^{-17}$), amplification would
require $\approx 3 \times 10^8$ repetitions; for 16 blocks ($p \approx 10^{-34}$),
$\approx 10^{17}$ - clearly impractical.
Mid-circuit measurement
reuse, deferred post-selection (accept only the final phrase's ancilla),
or a fundamentally different approach to phrase-level coherence will be
required for musically useful chain lengths.

\subsection{Circuit Synthesis Cost}
\label{sec:tech_circuit_synthesis}

The statevector implementation above applies a different chord vector
$|\mathbf{c}_i\rangle$ for each of the $N_\text{pairs}$ melody basis states,
controlled on the melody register.  In a real device implementation, this
controlled map would need to be synthesised as a single unitary acting on the
full 19-qubit register.  Under naive unitary synthesis, constructing this
unitary requires enumerating all $N_\text{pairs}$ chord vectors, incurring
$O(N_\text{pairs})$ classical compilation cost and potentially $O(N_\text{pairs})$
circuit depth - eliminating the $O(1)$ invocation advantage.

The Fourier oracle's QFT structure offers a path around this: a compact
circuit in which the chord register is populated via a QFT over pitch-class
space (parametrised by the $K$ Fourier coefficients) rather than by
explicit per-pair \texttt{StatePreparation}.  The circuit in
Section~\ref{sec:fourier_circuit_note} demonstrates this structure at
$C_\text{vocab} = 7$, achieving 375 gates without enumerating pairs.
However, that circuit encodes a \emph{single} note pair's chord vector, not
the full $N_\text{pairs}$-controlled map.  Extending it to a
melody-register-controlled oracle that applies the correct chord vector for
each basis state without $O(N_\text{pairs})$ circuit cost is an open circuit
design question.  The present paper establishes the architectural feasibility
in simulation; the efficient hardware-compilable oracle circuit is deferred
to future work.

\section{Limitations}
\label{sec:limitations}

\paragraph{Hardware availability.}
All results are from classical simulation; hardware requirements are detailed in
Section~\ref{sec:advantage_requirements}.

\paragraph{Post-selection compounding.}
The joint post-selection probability in a four-block chain is approximately
$4.5 \times 10^{-9}$, falling exponentially with chain length.  The H-chain
is not a scalable composition engine at the present simulation scale.

\paragraph{Scale.}
At the deliberately minimal scale described in Section~\ref{sec:architecture}, no
formal perceptual evaluation is reported.

\paragraph{Classical inter-block conditioning.}
Context passes classically between blocks, breaking full quantum coherence; a
monolithic four-note/four-chord single quantum event is a future hardware target.

\paragraph{No output quality advantage at current scale.}
At $N = 49$, the pipeline output is statistically indistinguishable from a
conditioned classical Markov chain baseline.  The value proposition is
architectural: the advantage would apply at large $N$, subject to the conditions
in Section~\ref{sec:advantage_requirements}.

\paragraph{Penalty parameters are author-specified.}
The Narmour+KK penalty values are not fitted to a corpus; the qualitative step
preference is robust to parameter variation (Section~\ref{sec:tech_sensitivity})
but corpus fitting remains an open task.

\paragraph{Leading-tone resolution.}
The Narmour+KK matrix encodes melodic \emph{interval implication} but not
\emph{harmonic tendency}: it has no rule that B must resolve upward to C when
functioning as a leading tone over V.  Consequently, the B$\to$C resolution
rate is only 21.3\% when V or vii$^\circ$ moves to I
(Section~\ref{sec:harmony_check}).  Context-sensitive diagonal penalties
(e.g.\ suppressing B$\to$non-C when $c_\text{prev} = \text{V}$) would close
this gap without changing the pipeline structure.  In common-practice writing
the leading tone may resolve in an inner voice, so this is a gap in the
single-voice encoding rather than a failure of functional harmony.

\paragraph{Circuit synthesis and the pipeline advantage claim.}
The Fourier oracle is currently implemented via statevector simulation; gate-level
equivalence was verified on a scaled-down test system (fidelity 1.0000,
Section~\ref{sec:oracle_impl}), so the open question is not correctness but
compilation cost.  Extending the Fourier circuit to a melody-register-controlled
oracle avoiding $O(N_\text{pairs})$ synthesis depth is an open question
(Section~\ref{sec:tech_circuit_synthesis}).

\section{Conclusion}

This work demonstrates that a coherent HHL+oracle pipeline - the mechanical
prerequisite for HHL's exponential speedup to be realised in musical generation -
is achievable at proof-of-concept scale.  The claim is not that quantum advantage
has been achieved, but that the architecture required for it has been constructed
and validated: HHL is coherently coupled to a Fourier harmonic oracle without
intermediate measurement, selecting both melody notes and a two-chord harmonic
progression simultaneously (as described above).
The musical examples - two melody notes over two block chords - are deliberately
minimal: the goal is to demonstrate that the coherent architecture works and that
the Fourier oracle produces measurably structured harmonic output consistent
with classical functional tonal grammar.

The coherent pipeline structure preserves the quantum coherence condition under
which the HHL speedup argument could apply at scale, subject to the requirements
in Section~\ref{sec:advantage_requirements}.
Quantitative results are in Tables~\ref{tab:oracle_comparison}-\ref{tab:k_chord_tone}
and Figures~\ref{fig:chain_k4}-\ref{fig:chain_k10}.

Future work: hardware implementation as fault-tolerant HHL becomes tractable;
extension to a monolithic four-note single quantum event as simulators improve;
exploration of additional oracle designs (e.g.\ walk-based or adaptive-$K$
oracles conditioned on melodic context, note-chord fit coupling via controlled
rotations); application to richer harmonic vocabularies beyond C major;
and extension of the independent harmony validation (Section~\ref{sec:harmony_check})
to larger generated sequences and richer chord vocabularies, including
corpus-frequency comparison against databases such as Hooktheory
(65{,}000+ annotated songs) to complement the rule-based checks reported here.

\bibliographystyle{apalike}

\end{document}